\documentclass[10pt,  conference]{IEEEtran}
\usepackage{todonotes}
\usepackage[nocompress]{cite}
\usepackage{url}
\usepackage{enumitem}
\usepackage{epstopdf}
\PrependGraphicsExtensions{.tif, .tiff}
\usepackage{graphicx}
\usepackage{caption}
\usepackage{amsmath}
\usepackage{tabu}
\usepackage{multirow}
\usepackage{adjustbox}
\usepackage{mathptmx}
\usepackage{amsfonts}
\usepackage{algpseudocode}
\usepackage{soul}
\usepackage[bookmarks=false]{hyperref}
\usepackage{footnote}
\usepackage{lipsum}
\usepackage[ruled,vlined, linesnumbered]{algorithm2e}
\usepackage{subcaption}
\usepackage{listings}
\usepackage{booktabs}
\usepackage{tikz}
\usetikzlibrary{positioning}
\soulregister\cite7
\soulregister\ref7
\soulregister\pageref7
\soulregister\secref7
\soulregister\tabref7

\definecolor{ao}{rgb}{0.0, 0.5, 0.0}
\definecolor{db}{rgb}{0.2, 0.2, 0.6}
\definecolor{cadmiumgreen}{rgb}{0.0, 0.42, 0.24}

\newcommand{\tabref}[1]{Table~\ref{#1}}

\newcommand{\secref}[1]{Section~\ref{#1}}

\SetAlFnt{\footnotesize}

\newcommand{\shrinkspace}{\vspace{-5mm}}

\begin{document}
\title{FedLesScan: Mitigating Stragglers in Serverless Federated Learning}
\author{\IEEEauthorblockN{Mohamed Elzohairy\IEEEauthorrefmark{1}, Mohak Chadha\IEEEauthorrefmark{1}, Anshul Jindal\IEEEauthorrefmark{1}, Andreas Grafberger\IEEEauthorrefmark{1}, Jianfeng Gu\IEEEauthorrefmark{1}, \\ Michael Gerndt\IEEEauthorrefmark{1}, Osama Abboud\IEEEauthorrefmark{2}\\}
\IEEEauthorblockA{\IEEEauthorrefmark{1}Chair of Computer Architecture and Parallel Systems, Technische Universit{\"a}t M{\"u}nchen \\
Garching (near Munich), Germany \\}
\IEEEauthorblockA{\IEEEauthorrefmark{2}Huawei Technologies, Munich, Germany \\} 
Email: \{mohamed.elzohairy, mohak.chadha, anshul.jindal, andreas.grafberger, jianfeng.gu\}@tum.de, \\gerndt@in.tum.de, osama.abboud@huawei.com}

\maketitle
\pagenumbering{gobble}


\begin{abstract}

Federated Learning (FL) is a machine learning paradigm that enables the training of a shared global model across distributed clients while keeping the training data local. While most prior work on designing systems for FL has focused on using stateful always running components, recent work has shown that components in an FL system can greatly benefit from the usage of serverless computing and Function-as-a-Service technologies. To this end, distributed training of models with serverless FL systems can be more resource-efficient and cheaper than conventional FL systems. However, serverless FL systems still suffer from the presence of \textit{stragglers}, i.e., slow clients due to their resource and statistical heterogeneity. While several strategies have been proposed for mitigating stragglers in FL, most methodologies do not account for the particular characteristics of serverless environments, i.e., cold-starts, performance variations, and the ephemeral stateless nature of the function instances. Towards this, we propose \emph{FedLesScan}, a novel  clustering-based semi-asynchronous training strategy, specifically tailored for serverless FL. \emph{FedLesScan} dynamically adapts to the behaviour of clients and minimizes the effect of stragglers on the overall system. We implement our strategy by extending an open-source serverless FL system called \emph{FedLess}. Moreover, we comprehensively evaluate our strategy using the 2$^{nd}$ generation Google Cloud Functions with four datasets and varying percentages of stragglers. Results from our experiments show that compared to other approaches \emph{FedLesScan} reduces training time and cost by an average of 8\% and 20\% respectively while utilizing clients better with an average increase in the effective update ratio of 17.75\%.

\end{abstract}



\section{Introduction}
\label{sec:intro}

The number of heterogeneous edge devices such as modern smartphones and IoT have significantly increased over the past few years. Due to the presence of several smart sensors, their increasing popularity, and computation power, these edge devices are able to accumulate and process enormous amounts of data each day~\cite{chiang2016fog}. Coupled with distributed machine learning (ML) techniques, the data generated by these devices can be used to solve challenging AI tasks~\cite{NvidiaClara2020, resnet}. With the increasing privacy concerns of the data holders and recent legislations on data protection and privacy such as the European General Data Protection Regulation (GDPR)~\cite{EUdataregulations2018}, Federated Learning (FL) has emerged as a novel distributed ML training paradigm that enables collaborative on-device training of ML models~\cite{mcmahan2017communication}. In contrast to the traditional centralized ML approach~\cite{lecun2015deep}, devices (\texttt{clients}) in FL never share their private data directly but learn a shared global model by optimizing its parameters on each device locally and sending back only the updated parameters. Following this, the local model updates are aggregated to form the shared global model. 

A traditional FL system consists of two main entities, i.e., \texttt{clients} and a \texttt{central server}. While most prior work on designing FL systems has relied on using stateful always running components~\cite{NvidiaClara2020, paddle, flower, fedml}, recent work~\cite{serverlessfl, fedless, jayaram2022lambda} has shown that both entities in a traditional FL system can greatly benefit from the usage of stateless serverless computing technologies, particularly Function-as-a-Service (FaaS)~\cite{cncf-serverless-whitepaper} ~(\S\ref{sec:faas}). FaaS is an emerging cloud-based programming paradigm that enables developers to focus on the application logic, while responsibilities such as infrastructure management, resource provisioning, and scaling are handled by the cloud service providers. In FaaS, the user implements fine-grained  functions that are executed in response to external triggers such as HTTP requests and deploys them into a FaaS platform such as Google Cloud Functions (GCF)~\cite{gcloud-functions-2}. In serverless FL, \texttt{clients} are independent functions deployed on a FaaS platform and capable of computing their model updates.




FaaS offers an attractive cloud model with advantages such as \texttt{scale-to-zero} for idle functions, a \texttt{pay-per-use} billing policy, and rapid automatic scaling on a burst of function invocation requests. Since a common practice in FL is to utilize only a small percentage of clients in a training round~\cite{bonawitz2019federated,mcmahan2017communication}, leaving the majority of clients idle until they are selected in a later round, using FaaS functions as FL clients can improve resource efficiency and reduce total training costs~\cite{fedless}. Moreover, due to the elasticity offered by FaaS platforms, using FaaS functions as aggregators in the FL server can improve aggregation latency, resource-efficiency, and significantly diminish aggregation costs~\cite{fedless, jayaram2022lambda, jayaram2022adaptive, jayaram2022just}. In addition, using FaaS technologies for FL can reduce the burden on the individual data holders to manage the complex infrastructure for their clients.


\begin{figure}[t]
\centering
\includegraphics[width=0.46\textwidth]{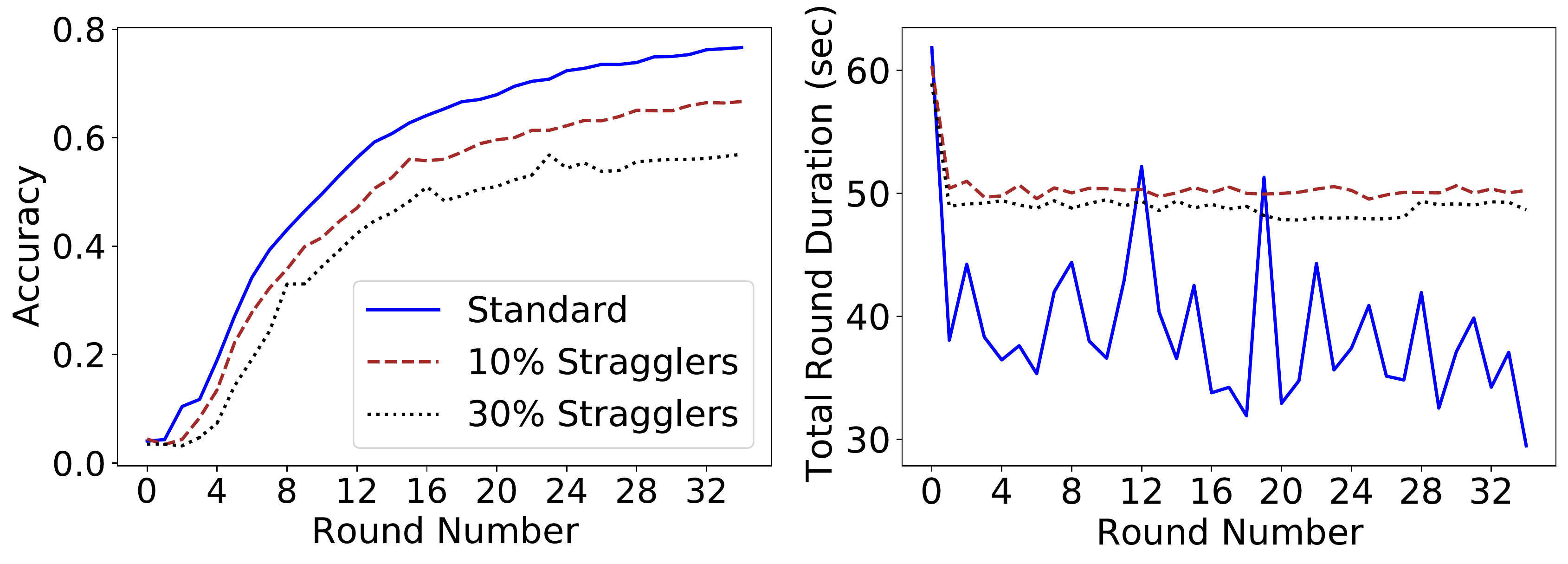}
\caption{Trained model accuracy (Left) and average FL training round duration (Right) for varying percentages of stragglers in the serverless FL system~\cite{fedless} for the Google Speech Dataset~\cite{speech} with the \texttt{FedAvg} algorithm~\cite{mcmahan2017communication}. All clients are deployed using the 2$^{nd}$ generation GCFs~\cite{gcloud-functions-2}  in the \texttt{europe-west-4} region.} 
\label{fig:motivation_example}
\shrinkspace
\end{figure}

Resource and statistical heterogeneity restrict the collaborative training process in large-scale FL systems~\cite{flmec}. Resource heterogeneity is caused by the difference in the computational and communication capabilities of the clients involved in the FL training process. On the other hand, statistical heterogeneity is caused by the presence of \textit{unbalanced} \textit{non-IID} data on clients in FL, i.e., different clients can have a different number of data records and one client's dataset is not representative of the full data distribution across all clients. As a result, the slower clients, i.e., \texttt{stragglers} greatly affect the global model training process. The usage of FaaS functions as FL clients greatly reduces the effect of stragglers on the overall training costs since client functions are only billed for their execution. However, the presence of stragglers in a serverless FL system can still reduce the accuracy of the trained global model and significantly increase the total FL round duration as shown in Figure~\ref{fig:motivation_example}. Note that, due to the synchronous nature of the \texttt{FedAvg} algorithm~\cite{mcmahan2017communication}, the FL server waits for all clients to send back their updates or time-out leading to relatively constant round durations.

To mitigate stragglers in FL, several synchronous~\cite{fedprox, gradientreduced} and asynchronous techniques~\cite{fed_async, asofed} have been proposed. Synchronous strategies have the disadvantages of increased training time and resource utilization, while asynchronous strategies suffer from higher communication costs and often require a persistent communication link between the FL server and the clients. Moreover, most prior strategies~\cite{fedAt, csafl, safa, fleet, fedskel} do not account for the particular characteristics of serverless environments, i.e., cold-starts, performance variations~\cite{wang2018peeking}
, and the ephemeral stateless nature of the function instances (\S\ref{sec:faas}). To this end, this paper advances the state-of-the-art in Serverless FL by proposing \texttt{FedLesScan}, a novel clustering-based semi-asynchronous training strategy, specifically tailored for serverless FL. Our strategy consists of two main components. First, an adaptive clustering-based client selection algorithm that selects a subset of clients for training based on their previous behavior. Second, a staleness-aware aggregation scheme to mitigate slow model updates and avoid wasted contribution of clients. 

Towards efficient serverless FL, our key contributions are:
\begin{itemize}
    \item We propose \texttt{FedLesScan}, a novel training strategy designed for serverless FL.
    \item We implement \texttt{FedLesScan} by extending an open-source serverless FL system called \emph{FedLess}~\cite{fedless}. This represents a real system that can be used by data holders for FL. Our implementation can be found here\footnote{\url{https://tinyurl.com/3utdhuuu}}.
    \item We comprehensively evaluate our methodology using the 2$^{nd}$ generation Google Cloud Functions~\cite{gcloud-functions-2} for upto 200 concurrent clients on four different datasets against two other popular FL training approaches wrt  accuracy, round utilization, training time, and cost. We demonstrate that \texttt{FedLesScan} constantly outperforms the other strategies wrt the different metrics for varying number of stragglers in the system.
\end{itemize}

The rest of the paper is structured as follows. \S\ref{sec:faas} gives a brief overview on FaaS. In \S\ref{sec:relatedWork}, prior work on serverless FL, stragglers in FL and serverless FL is described. \S\ref{sec:sysdesign} describes our extensions to \emph{FedLess} to make it more easier to use and for implementing our strategy. In \S\ref{sec:algo}, we describe \texttt{FedLesScan} in detail. \S\ref{sec:results} describes our experimental results. Finally, \S\ref{sec:conclusion} concludes the paper and presents an outlook.

\section{Function-as-a-Service}
\label{sec:faas}
FaaS is a key enabler of the serverless computing paradigm and has gained significant popularity and widespread adoption in various application domains such as high performance computing~\cite{hep, chadha2021architecture}, machine learning~\cite{Carreira2019}, edge computing~\cite{tinyfaas, fado}, and heterogeneous computing~\cite{courier, fncapacitor}. It enables a traditional monolithic application to be decomposed into fine-grained functions that are packaged as independent containers~\cite{docker} and uploaded to a FaaS platform. Examples of commercial and open-source FaaS platforms include Google Cloud Functions~\cite{gcloud-functions-2}, AWS Lambda~\cite{aws-lambda}, and OpenFaaS~\cite{openfaas} respectively. The functions are invoked on events such as HTTP, \texttt{gRPC} requests, or insertion of data into a database. On function invocation, the FaaS platform creates an execution environment, i.e., \textit{function instance} which provides a secure and isolated language-specific \textit{runtime} environment for the function. When a function is invoked for the first time, a new function instance is created by the FaaS platform, i.e., cold start. Function cold starts can significantly increase function request latencies and are a major challenge in FaaS~\cite{agile, vhive, demystifying, tppfaas, slam}. The FaaS platform is also responsible for the on-demand autoscaling of function instances based on the number of events. Moreover, if there are no function invocation requests within a given time frame, the FaaS platforms automatically scales down the number of idle function instances to zero. Most commercial FaaS providers limit the maximum amount of memory that can be allocated to a function and its execution time. For instance, with  2$^{nd}$ generation GCFs~\cite{gcloud-functions-2}, these limits are $60$ minutes and $16$GB, respectively. Moreover, none of the current FaaS offerings support execution of functions on GPUs. However, due to the rapid development in FaaS offerings from commerical cloud providers, these limitations might disappear or be relaxed soon~\cite{Fox2017}.

\section{Related Work}
\label{sec:relatedWork}

\subsection{Serverless Federated Learning}
\label{sec:servfl}
While using serverless technologies for distributed ML training has been researched extensively in the literature~\cite{Wang2019a, Carreira2019, Jiang2021, mlless, ali2022smlt}, exploring the use of FaaS functions for FL~\cite{fedless, serverlessfl, jayaram2022lambda} is a relatively new research direction. The first work in this domain was by Chadha et al.~\cite{serverlessfl}, in which they proposed \emph{FedKeeper}, a tool for orchestrating the training of Deep Neural Network (DNN) models using FL for clients distributed across a combination of heterogeneous FaaS platforms. As an evolution of \emph{FedKeeper}, Grafberger et al.~\cite{fedless} present \emph{FedLess}, a system and framework for scalable FL using serverless computing technologies.
\emph{FedLess} is cloud-agnostic, supports all major commercial and open-source FaaS platforms (\S\ref{sec:faas}), and enables training of arbitrary DNN models using the Tensorflow~\cite{tensorflow} library. Furthermore, it includes several important security features such as authentication/authorization of client functions using AWS Cognito~\cite{aws-cognito} and supports privacy-preserving FL training of models using Differential Privacy~\cite{Mothukuri2021,Kim2021,Wei2020}. Moreover, it was designed from the ground up for serverless environments and includes several performance optimizations such as global namespace caching, running average model aggregation, and federated evaluation. A more recent work by Jayaram et al.~\cite{jayaram2022lambda}, proposes an adaptive aggregation methodology for FL systems based on FaaS functions over multiple steps. They prototype their solution using Ray~\cite{ray} and show a significant reduction in resource utilization and costs as compared to the conventional stateful aggregation schemes. However, their proposed system does not support serverless FL training of DNN models. As a result, in this paper, we implement and evaluate our strategy using \emph{FedLess}~\cite{fedless}. 


\subsection{Stragglers in Federated Learning}
\label{sec:stragglersfl}

The different proposed approaches for straggler mitigation in FL can be divided into three categories, i.e., \textit{synchronous}~\cite{fedprox}, \textit{asynchronous}~\cite{asofed,fed_async}, and \textit{semi-asynchronous}~\cite{fedAt, safa, csafl}. One of the most popular synchronous approaches is \texttt{FedProx}~\cite{fedprox}, which is based on \texttt{FedAvg}~\cite{mcmahan2017communication} with two minor differences. First, a custom loss function for client-side training that limits the varying effects of local model updates. Second, toleration for partial work, i.e., clients can perform a variable amount of work to accommodate constraints in terms of hardware and network. However, incorporating partial work requires tailoring the number of local epochs for each client individually, which might be infeasible for a significantly large number of clients. Moreover, in contrast to \texttt{FedLesScan} it relies on random selection of clients which makes it sensitive to stragglers. In~\cite{fed_async}, Xie et al. propose an asynchronous federated optimization algorithm called \texttt{FedAsync} that relies on a parameter server architecture to invoke and synchronize FL clients. The server consists of two parallel threads, i.e., a \textit{scheduler} and an \textit{updater}. The scheduler thread periodically triggers the FL clients to perform training using the latest global model, while the updater thread receives the client model updates and directly aggregates them to the global
model. However, in the context of serverless, this approach has significantly higher resource and communication costs (\S\ref{sec:faas}, \S\ref{sec:servfl}) since it would require a function invocation to perform aggregation with the global model after each individual client update.



Semi-asynchronous approaches aim for the resilience of asynchronous strategies while decreasing communication costs. Strategies such as \texttt{CSAFL}~\cite{csafl} and \texttt{FedAt}~\cite{fedAt} use a grouping mechanism to partition clients into a specific number of groups based on their training times. However, a limitation of this approach is that the number of groups cannot be adapted dynamically during training based on the behaviour of clients. Other strategies, such as \texttt{SAFA}~\cite{safa} rely on invoking all clients in every round, while using the fastest responses and caching the rest for subsequent rounds. Similar to \texttt{FedAsync}, this approach requires a thread to wait for asynchronous client updates. Furthermore, with \texttt{SAFA} the communication costs are significantly increased. In contrast, \texttt{FedLesScan} provides a clustering mechanism that adjusts the number of client groups dynamically during training. Moreover, we intelligently select a fixed number of clients in each round and use a staleness-aware aggregation scheme that does not require waiting for the asynchronous client updates.





\begin{figure}[t]
  \centering
   \includegraphics[width=0.40\textwidth]{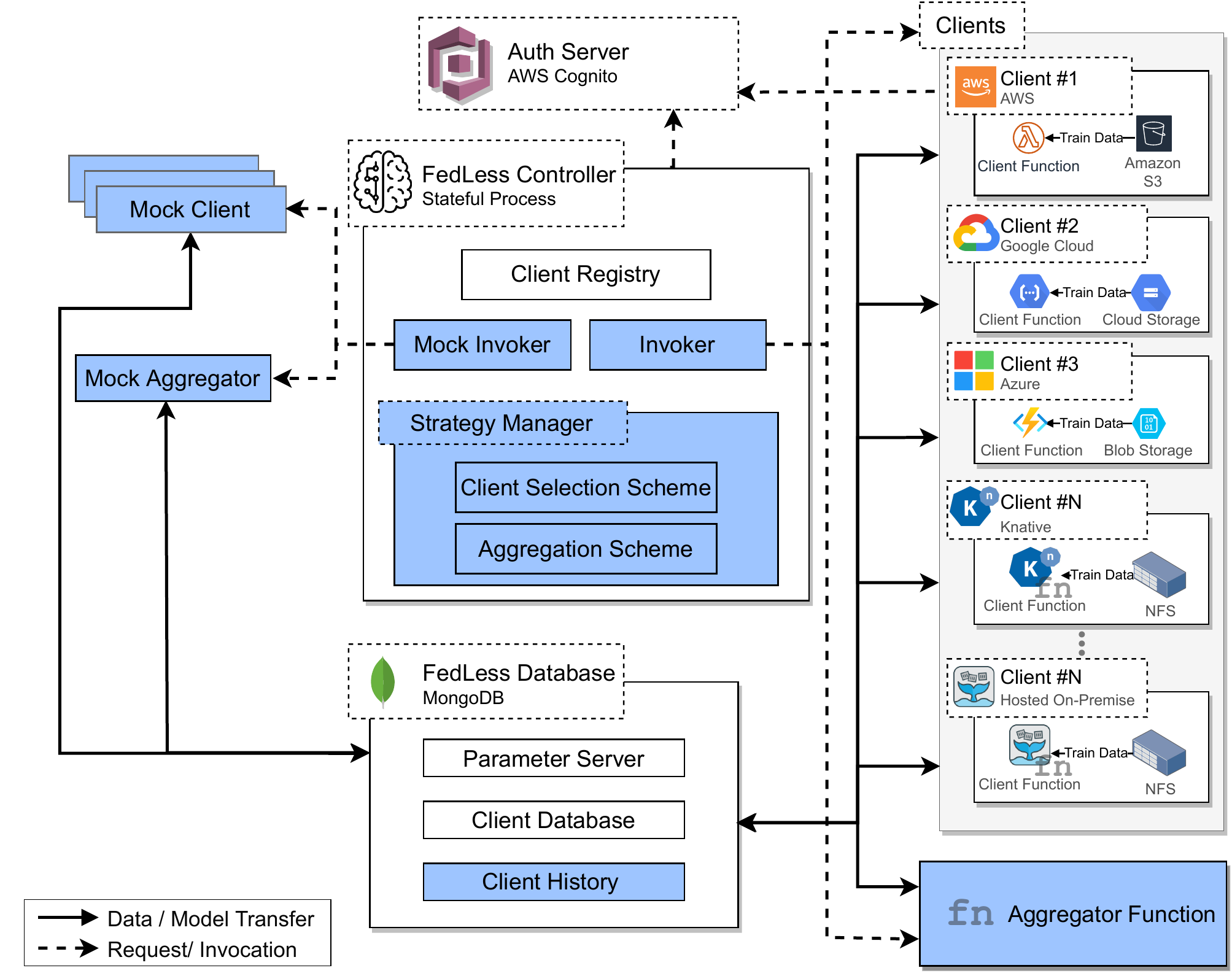}
  \caption{Modified architecture of \emph{FedLess}~\cite{fedless}. The components highlighted in blue are our extensions and modifications to the system.}\label{fig:arc-fedless}
\shrinkspace
\end{figure}

\subsection{Stragglers in Severless Federated Learning}
\label{sec:stragglersservfl}

Despite several advantages of the FaaS model, such as elasticity, resource-efficiency, and costs, such systems have weak reliability guarantees~\cite{serverless_reliability, serverless_reliability_2}. Node failures can cause requests to be dropped or even executed multiple times. For instance, the Service Level Objective (SLO) for GCF is an uptime percentage of 99.95\%~\cite{gcf_sla} as compared to 99.99\% for multi-zone compute instances~\cite{gcp_compute}. As a result, FL clients hosted on a FaaS platform can frequently fail. Furthermore, with \texttt{scale-to-zero} (\S\ref{sec:faas}), the FL clients might frequently undergo cold starts which can significantly impact the FL training round duration. Moreover, with most commercial FaaS platforms, the FL client function instances are launched on the FaaS platform's automatically provisioned virtual machine (VM) offerings. However, the user is not aware of the details of the provisioned VMs such as the CPU architecture. To this end, the performance of the FL client functions can significantly vary depending on the characteristics of the provisioned VM~\cite{behind, wang2018peeking, chadha2021architecture, mohanty2018evaluation}. To the best of our knowledge, \texttt{FedLesScan} is the first strategy designed to facilitate efficient FL in serverless environments.

\section{System Design}
\label{sec:sysdesign}
In this section, we describe our extensions to the open-source system and framework \emph{FedLess}~\cite{fedless}. 



\subsection{Extensions to FedLess}
\label{sec:extfedless}
Figure~\ref{fig:arc-fedless} provides an overview of the modified system architecture of \emph{FedLess}~\cite{fedless}, with the different added system components highlighted in blue. 
The first significant enhancement is to the \emph{FedLess} controller. Previously, the controller required a complex deployment configuration and ran on a Kubernetes (K8s)~\cite{k8s} cluster with deployed OpenWhisk (OW)~\cite{openwhisk}. To overcome this, we completely removed the dependencies on  OW and K8s. To this end, the controller is now a lightweight process that can run on any system that includes its dependencies without any infrastructure management. A major challenge with distributed systems is the cost~\cite{jonas2017occupy}, ease of development, and debugging. Previously, the development and debugging of client/aggregation functions in \emph{FedLess} required their deployment on an actual FaaS platform (\S\ref{sec:faas}). To mitigate this, we implemented a mocking system in \emph{FedLess} that enables developers to run and debug the entire system on a single machine. We achieved this by adding different mock components to the controller that simulate the behavior of the actual components. It should be noted that using mock clients and aggregator does not require any additional effort from the user's side and everything is handled internally by the Mock \textit{Invoker}. The mocking system is activated by passing the \texttt{-mock} flag to the controller. Moreover, we abstracted the aggregation support to enable its implementation with any FaaS platform.


To implement our strategy with \emph{FedLess}, we added a \textit{Strategy Manager} component to the controller. It is responsible for controlling the behavior of the selected strategy and contains two sub-components, i.e., the \textit{client selection} and the \textit{aggregation} scheme. The former is responsible for selecting the clients involved in a particular training round, while the latter is responsible for the type of aggregation algorithm used. Furthermore, to store the behavioral data of the clients, such as the number of failures, training duration, and the training rounds missed (\S\ref{sec:behavedata}), we added a client history collection to the database. The overall FL training workflow with our strategy remains similar to the one described in~\cite{fedless} with minor changes required for fetching/updating the behavioral data of clients.

\section{FedLesScan}
\label{sec:algo}
In this section, we describe our methodology for mitigating stragglers in serverless FL. First, we describe our approach for partitioning clients into separate tiers. Following this, we describe the behavioral data we collect for the different clients. After this, we describe our clustering-based client selection algorithm. Finally, we describe our staleness-aware aggregation strategy. 



\begin{algorithm}[t]

\SetAlgoLined
\SetKwFunction{train}{Client\_Update}
\SetKwFunction{fitRound}{Train\_Global\_Model}
\SetKwFunction{startTime}{Start\_Timer}
\SetKwFunction{stopTime}{Stop\_Timer}
\SetKwProg{Pn}{Function}{:}{\KwRet}
\DontPrintSemicolon
\textbf{Fedless Controller:}
\;

\Pn{\fitRound{$Clients$, $round$}}
{   
    $selected$ = Select\_Clients ($clients$, $round$, $maxRounds$, $clientsperRound$)\;
    $success$, $failures$ = Invoke\_Clients ($selected$)\;
    \For{each $client$ in $success$}
    {
        Set cooldown of $client$ to zero.\;
        Update $client$ history.\;
    }
    \For{each $client$ in $failures$}
    {
        Add current $round$ to the $clients$ missed rounds.\;
        Update cooldown of the $client$ according to Eq.~\ref{eq: cooldown}.\;
        Update $client$ history.\;
    }
}

\textbf{Fedless Client:}
\;
\Pn{\train{$hyperParameters$, $round$}}
{
    
    Load client history.\;
    \startTime{}\;
    Load model and dataset.\;
    Train model\;
    \stopTime{}\;
    Save updated model to database.\;
    Add measured time to $client$ history.\;
    \If{$round$ present as missed round in $client$ history}{
        Remove missed round from $client$ history.\;
    }
    Update $client$ history.\;
}
\caption{Modifed \texttt{FedLess}~\cite{fedless} controller and client routines.}
\label{alg:client-side}

\end{algorithm}

\subsection{Partitioning Clients}
\label{sec:tiers}
Our selection strategy separates reliable clients that do not miss their training rounds due to slow updates, timeouts, or failures from stragglers. Towards this, we partition the clients into three groups, i.e., (i) \textit{rookies}, (ii) \textit{participants}, and (iii) \textit{stragglers}. \textit{Rookies} are clients which have not participated in the FL training process and for which no behavioral data exists (\S\ref{sec:behavedata}). \textit{Participants} are the group of clients that can participate in the clustering process (\S\ref{sec:behavedata}, \S\ref{sec:ClientSelection}). In our strategy, we use clustering to group reliable clients with similar behavior together. Finally, \textit{stragglers} are clients that have missed one or more consecutive training rounds. They have the lowest priority in our selection process (\S\ref{sec:ClientSelection}) and are characterized by using a variable called \texttt{cooldown} as described in \S\ref{sec:behavedata}. Note that, our client selection strategy (\S\ref{sec:ClientSelection}) adapts to the behavior of the clients as the FL training progresses. As a result, tier-2 clients can move to tier-3 and vice-versa.

\subsection{Gathering Behavioral Data}
\label{sec:behavedata}
Our client selection strategy (\S\ref{sec:ClientSelection}) depends on the data collected from the clients' behavior in previous training rounds. Towards this, for each client we collect three attributes, i.e., \textit{training time}, \textit{missed rounds}, and \textit{cooldown}. \textit{Training time} is the time taken by the client to complete local model training. \textit{Missed rounds} is a list that contains the round numbers of the rounds missed by a client. We use this to calculate a client's penalty as described in \S\ref{sec:ClientSelection}. \textit{Cooldown} represents the number of rounds a client has to stay in the last tier and cannot participate in clustering (\S\ref{sec:tiers}). We evaluate the cooldown period from the client’s last missed round using Equation~\ref{eq: cooldown}. For instance, if a client missed round $2$, the cooldown is set to $1$. Moreover, if the same client missed round $4$, the cooldown is multiplied by two. As a result, the clients with \textit{cooldown} greater than zero are characterized as \textit{stragglers}. Towards this, using \textit{cooldown} can reduce the impact of temporarily slow or unavailable clients by lowering their priority for a certain number of rounds (\S\ref{sec:tiers}, \S\ref{sec:ClientSelection}).

\begin{footnotesize}
\begin{equation}
    cooldown = \left\{
     \begin{array}{@{}l@{\thinspace}l}
      0  &: \text{if the client completed training in time}\\
      1 &: \text{if cooldown = zero }\\
      cooldown \times 2 &: \text{otherwise}\\
     \end{array}
  \right.  
  \label{eq: cooldown}
\end{equation}
\end{footnotesize}

To gather and update the different client attributes, we modify the \emph{FedLess} controller (\S\ref{sec:sysdesign}) and FL client routines as shown in Algorithm~\ref{alg:client-side}. In each FL training round, the controller runs the \texttt{Train\_Global\_Model} routine, while the clients run the \texttt{Client\_U\-pdate} routine to locally train the model. \textit{nClientsPerRound} represents the maximum number of clients that must be selected in every round and \textit{maxRounds} represents the maximum number of allowed training rounds. 

Initially, the controller selects a subset of clients, invokes them, and then waits until they finish or a timeout occurs (Lines 3-4). Following this, we iterate over each successful client response and reset the \textit{cooldown} variable (Lines 5-8). Since the controller doesn't know if the client was slow or crashed, it assumes that the remaining invoked clients failed to finish. Subsequently, we update their \textit{missed rounds} and \textit{cooldown} attributes (Lines 9-12). At the client-side, we first load its behavioral history from previous rounds (Line 17). Following this, we measure the time for the client to load the global model, its local dataset, and train the model with its local data (Lines 18-21). Following this, each client sends its local model updates to the database (Line 22) and updates its training time for the current round (Line 23). Furthermore, slower clients that finished a round later can correct information about their missed rounds. As described before, the controller considers clients that did not finish the round in time as crashed. Therefore, distinguishing between crashed and slow clients is done on the client side. This is done by deleting the current round from their missed rounds list (Lines 24-26). Finally, the client updates its information in the database (Line 27).

\begin{footnotesize}
\begin{equation}
  totalEma = trainingEma + missedRoundEma \times maxTrainingTime
 \label{eq:total-ema}
\end{equation}
\end{footnotesize}

\begin{algorithm}[htp]
\SetAlgoLined
\SetKwFunction{selectClients}{Select\_Clients}
\SetKwProg{Pn}{Function}{:}{\Return{[$rookies$ + $cluster\_clients$ + $straggler\_clients$]}}
\DontPrintSemicolon

\Pn{\selectClients{$clients$, $round$, $maxRounds$, $clientsperRound$}}
{

    Characterize clients as $rookies$, $participants$, and $stragglers$.\;
    \If{ \#$rookies$ $\geq$ $clientsperRound$}{
        \Return{Randomly sample $clientsperRound$ from $rookies$.}
    }
    Calculate \#$cluster\_clients$ required from \textit{participants}.\;
    Calculate \#$straggler\_clients$ required from \textit{stragglers}.\;
    Randomly sample \#$straggler\_clients$ from \textit{stragglers}.\;
    $clusteringData$ = []\;
    \For{each client in $participants$}
    {
        Calculate EMA of $client$ based on its training time.\;
        Calculate missed round ratios and $missedRoundEma$\;
        Update $clusteringData$.\;
    }
    Obtain cluster labels using DBScan and $clusteringData$.\;
    Sort Clusters.\;
    Sample \#$cluster\_clients$ from sorted clusters.\;
}

\caption{Client selection}
\label{alg:client-selection}

\end{algorithm}

\subsection{Selecting Clients}
\label{sec:ClientSelection}

Algorithm~\ref{alg:client-selection} describes our strategy for client selection in a particular FL training round. Our strategy promotes fair selection for reliable clients while involving stragglers less in the training process. Initially, we characterize the clients as \textit{rookies}, \textit{participants}, or \textit{stragglers} (\S\ref{sec:tiers}) based on their previous behaviour (Line 2). First, we randomly select the required number of clients in a round from the pool of \textit{rookie} clients. If the number of available \textit{rookie} clients is greater than the number of required clients, the algorithm terminates (Line 3-5). If not, we calculate the number of clients required from the \textit{participants} and the \textit{stragglers} (Lines 6-7). Note that, clients from \textit{stragglers} are only selected if the number of clients selected from the first and second tiers is not sufficient. Following this, we randomly sample the required number of straggler clients from the \textit{stragglers} (Line 8).

The reason for prioritizing rookies first is to guarantee that every client gets a chance to contribute to the global model. Moreover, it provides data on clients' behavior which is used to cluster clients in future rounds. Note that the number of rookies decreases as the training progress, until it reaches zero. In this case, the algorithm fully relies on the dynamic clustering of tier two clients.


For the clients that will participate in clustering, we calculate two features. First, \textit{trainingEma} which represents an exponential moving average (EMA)~\cite{ema} on the previously recorded training times (Line 11). We use EMA since a weighted average better represents the current behavior of the client by giving higher weight to the recently recorded times. Second, \textit{missedRoundEma} which is a penalty factor based on the previous rounds missed by the client ~(\S\ref{sec:behavedata}). This satisfies two objectives. First, recent failures should have higher penalties. Second, the penalty should decrease as the training progresses if the client becomes more reliable. To compute \textit{missedRoundEma}, we divide the round numbers in the \textit{missed rounds} list by the current round number to get a list of ratios. Following this, we take an EMA on the computed ratios (Line 12). As the training progresses, the effect of a specific missed round decreases because the current round number increases. The computed features are added to the data required for clustering (Line 13). Following this, we use the collected data as input to the  \texttt{DBSCAN}~\cite{dbscan} algorithm for partitioning it into separate clusters, each with a specific label (Line 15). For generating clusters, \texttt{DBSCAN} relies on an $\epsilon$ parameter that represents the maximum distance between two samples to be considered in the neighborhood of each other. For simplicity, we treat outliers as a single cluster. To find a value for $\epsilon$, we do a grid search and select the parameter value that yields the highest \textit{Calinski-Harabasz index}~\cite{ch-index}. This score measures the ratio between intra-cluster and inter-cluster dispersion to evaluate the quality of the clusters. We chose this index because it is fast to compute and will not impact our running time. Moreover, due to the low time complexity of \texttt{DBSCAN}, i.e., \texttt{O(Nlog(N))} the time for computing clusters multiple times is insignificant compared to the overall round time. Following this, we sort the clusters according to the increasing order of the average \textit{totalEMA} of their members (Line 16). We calculate the \textit{totalEMA} using Equation~\ref{eq:total-ema}. For sampling clients from the sorted clusters, we start by first choosing the clients belonging to the faster clusters and gradually move to the slower clusters in the sorted list. Moreover, to provide maximum information to the global model and avoid sampling from the same cluster in every training round, we start choosing from the cluster corresponding to our current
training progress. This is determined by using the ratio between the
current round and the maximum number of rounds. We continue sampling from the clusters until we reach the required number of clients or there are no more clients to choose from. Finally, our algorithm returns a list of clients selected from \textit{rookies}, clusters, and \textit{stragglers}. 

\texttt{FedlesScan} accounts for offline clients and cold starts by not only relying on training time in the selection, but also by utilizing a dynamic penalty factor that accounts for delays and cold starts if they cause clients to miss their rounds. Furthermore, our strategy does not need any further computation on the client side, requiring clients to only be active during training making it suitable for serverless environments~(\S\ref{sec:stragglersfl}). 

\begin{footnotesize}
\begin{equation}
 w_{t+1} \longleftarrow \sum_{k=1}^{K} \frac{t_k}{t} \times \frac{n_{k}}{n} w^k_{t_k} 
 \label{eq:fedless-aggr}
\end{equation}
\end{footnotesize}

\subsection{Staleness-aware Aggregation}
\label{sec:stalelessaware}
Although our intelligent client selection strategy (\S\ref{sec:ClientSelection}) improves the efficiency of the system, stragglers are not completely eliminated. Stragglers might push their local model updates to the parameter server after the completion of an FL training round. Moreover, these updates might contain valuable information that can improve the performance of the global model. Towards this, we aggregate the delayed updates of the clients with a dampening effect asynchronously, i.e., delayed updates are considered the next time the aggregation function is called. 
Equation~\ref{eq:fedless-aggr} shows our updated aggregation function used to include stale updates. $w^k_{t_k}$ is the local model of client \textit{k} at round $t_k$ and $w_{t+1}$ is the global model after aggregation at round $t$. Furthermore, $n_k$ represents the cardinality of the dataset at client $k$ while $n$ is the total cardinality of the aggregated clients.  If the updates arrive at the same round ($t_k = t$), the equation becomes similar to \texttt{FedAvg}~\cite{mcmahan2017communication}. On the other hand, older updates ($t_k < t$), are dampened by $\frac{t_k}{t}$. To avoid obsolete updates from affecting the training, the aggregator uses a parameter $\tau$ to dictate the maximum age of updates included in the aggregation. Updates with $t - t_k \ge \tau$ are discarded by the aggregator. In our experiments, we use a value of two for $\tau$.
\section{Experimental Results}
\label{sec:results}
In this section, we first describe the datasets and model architectures used for evaluating our strategy. Following this, we describe the configuration for \emph{FedLess} and the FaaS client functions. Finally, we evaluate the performance of our strategy as compared to \texttt{FedAvg}~\cite{mcmahan2017communication} and \texttt{FedProx}~\cite{fedprox} wrt several metrics for varying number of stragglers. We chose these strategies since they are robust and work well in a serverless setting. For all our experiments, we follow best practices while reporting results and repeat them three times~\cite{8758926}. More extensive experimental results can be found here~\cite{thesismohamed}.


\subsection{Experiment Setup}

\subsubsection{Datasets}
\label{sec:datasets}
For our experiments, we use four datasets from different domains. These include image classification, i.e., (\textit{MNIST}~\cite{lecun-mnist}, \textit{FEMNIST}~\cite{cohen2017emnist}), language modeling, i.e., (\textit{Shakespeare}~\cite{shakespeare-gutenberg}), and speech recognition, i.e.,  (\textit{Google Speech}~\cite{speech}).

The MNIST handwritten Image Database consists of 60,000 images for training and 10,000 images for central evaluation. Similar to~\cite{fedless,mcmahan2017communication}, for simulating a non-IID setting, we sort the images by label, split them into 300 shards of 200 images each, and distribute the shards to the clients. We choose the FEMNIST and the Shakespeare datasets from a benchmarking framework for FL called \texttt{LEAF}~\cite{caldas2018leaf}. The FEMNIST dataset contains more than eight hundred thousand images, with an average of 226 images per client partition. The Shakespeare dataset contains sentences from \textit{The Complete Works of William Shakespeare}~\cite{shakespeare-gutenberg}. It is divided such that each role in each play is mapped to a specific partition. With more than four million samples, each of length 80 characters, a client is responsible for an average of 3743 samples. For this dataset, the task is to predict the next character in a sentence given the previous 80. We chose the real-world Google Speech dataset from the \texttt{FedScale}~\cite{fedscale} benchmark suite. This dataset was designed to build basic and helpful voice interfaces for applications with common words such as "Yes", "No," and directions. The primary goal of the speech recognition task is to train and evaluate keyword spotting systems to detect when a specific keyword is spoken~\cite{speech}. The dataset contains 105,000 samples of 1-second audio files distributed across 2618 clients. Out of the 2618 clients, $2168$ clients are for training, while the rest are for validation and testing. However, in our experiments we scale down the number of clients by four by using a custom mapping algorithm. As a result, we used 542 clients for the Google Speech dataset with each client containing training data of four corresponding clients from \texttt{FedScale}.

\subsubsection{Model Architectures and Parameters}
\label{sec:models}
For each dataset, we use a DNN model architecture suited for the particular task. For the MNIST, FEMNIST, and the Shakespeare datasets, we use the same model architecture used in ~\texttt{LEAF}~\cite{caldas2018leaf}. On the other hand, for the Google Speech dataset we use a relatively simpler model as compared to the one described in \texttt{FedScale}~\cite{fedscale}. However, with our model we achieve similar results for accuracy as compared to~\cite{fedscale}.

For MNIST, we use a 2-layer Convolutional Neural Network (CNN) with a 5x5 kernel. Each convolutional layer is followed by a 2x2 max-pooling layer.  The model ends with a fully-connected layer with 512 neurons and a ten-neuron output layer. Similarly, for FEMNIST we use a 2-layer CNN with a 5x5 kernel in which each convolutional layer is followed by a 2x2 max-pooling layer. The CNN layers are followed by a fully-connected layer with 2048 neurons, which is followed by an output layer with 62 neurons. For the Shakespeare dataset, we use a Long Short Term Memory (LSTM) recurrent neural network~\cite{hochreiter1997long}. The model consists of an embedding layer of size eight followed by two LSTM layers with 256 units and an output layer of size 82. Our model architecture for Google Speech consists of two identical blocks followed by an average pooling layer and an output layer with 35 neurons. A block contains two convolutional layers with a 3x3 kernel followed by a max-pooling layer. A dropout layer follows the max-pooling layer with a rate of \texttt{0.25} to avoid overfitting. For MNIST, FEMNIST, and Google Speech, we use Adam~\cite{adam} as the optimizer with a learning rate of $1e-3$. On the other hand, for Shakespeare, we use SGD~\cite{sgd} with a learning rate of $0.8$. The clients for the MNIST, FEMNIST, and Speech datasets train for five local epochs with a batch size of $10$, $10$, and five respectively. Due to the limitations of the current commercial FaaS platforms (\S\ref{sec:faas}), the clients for Shakespeare train for one local epoch with a batch size of $32$. The hyperparameters for the different datasets are shown in Table~\ref{tab:experiment-params}.

\begin{table}[t]
\centering
\begin{adjustbox}{width=\columnwidth,  center}
\begin{tabular}{|c|ccc|cc|}
\hline
\multirow{2}{*}{\textbf{Dataset}} & \multicolumn{3}{c|}{\textbf{Hyperparameters}}                                                           & \multicolumn{2}{c|}{\textbf{Number of Training Rounds}}                             \\ \cline{2-6} 
                                  & \multicolumn{1}{c|}{\textbf{Epochs}} & \multicolumn{1}{c|}{\textbf{Batch Size}} & \textbf{Learning Rate} & \multicolumn{1}{c|}{\textbf{Standard}} & \textbf{Straggler (\%)} \\ \hline
MNIST                             & \multicolumn{1}{c|}{5}               & \multicolumn{1}{c|}{10}                  & $1e-3$                 & \multicolumn{1}{c|}{60}                & 60                      \\ \hline
FEMNIST                           & \multicolumn{1}{c|}{5}               & \multicolumn{1}{c|}{10}                  & $1e-3$                  & \multicolumn{1}{c|}{40}                & 40                      \\ \hline
Shakespeare                       & \multicolumn{1}{c|}{1}               & \multicolumn{1}{c|}{32}                  & $0.8$                 & \multicolumn{1}{c|}{25}                & 25                      \\ \hline
Speech Command                    & \multicolumn{1}{c|}{5}               & \multicolumn{1}{c|}{5}                   & $1e-3$                  & \multicolumn{1}{c|}{35}                & 60                      \\ \hline
\end{tabular}
\end{adjustbox}
\caption{Experiment hyperparameters for the different datasets (\S\ref{sec:datasets}) and experiment scenarios (\S\ref{sec:scenarios}).}

\shrinkspace
\label{tab:experiment-params}
\end{table}

\subsubsection{Experiment configuration}
\label{sec:expconfig}
To scale our experiments, we deployed \emph{FedLess} (\S\ref{sec:sysdesign}) on a VM hosted by the LRZ compute cloud~\cite{lrzcc}. The VM was configured with $40$ virtual CPUs (vCPUs) and $177$GB of RAM. For the aggregator function, we use a self-hosted OpenFaaS~\cite{openfaas} cluster deployed on a VM with 10vCPUs and 45GB of RAM. We limit the memory of the aggregation function to $7$GB. Furthermore, for hosting our datasets we use a nginx~\cite{nginx} store running on a VM with 10vCPUs and 45GB of RAM.

For all our experiments, we deployed the FaaS based FL clients on the 2$^{nd}$ generation GCFs~\cite{gcloud-functions-2} based on Knative~\cite{knative}. Each client function had a memory limit of 2048MB and a timeout of 540 sec. For the MNIST, FEMNIST, and Google Speech datasets, we use $200/300$, $175/300$, and $200/542$ concurrent clients per round respectively. On the other hand, for Shakespeare, we use $50/100$ clients per round.

\begin{figure*}
\begin{subfigure}{\textwidth}
\centering
\includegraphics[width=\textwidth]{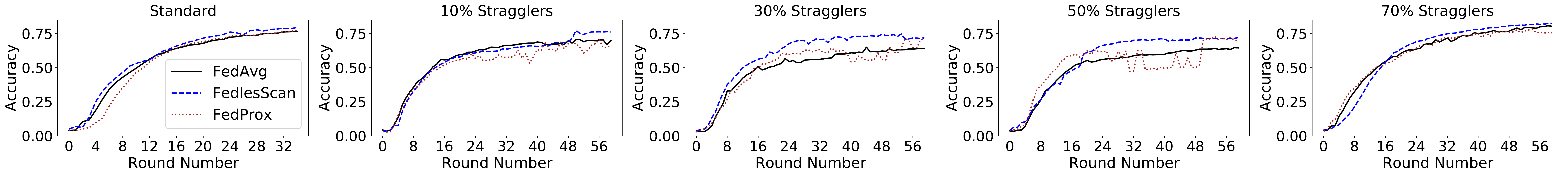}
\caption{Comparing Accuracy.}
\label{fig:acc-speech}
\end{subfigure}%
 \hfill

\begin{subfigure}{\textwidth}
\centering
\includegraphics[width=\textwidth]{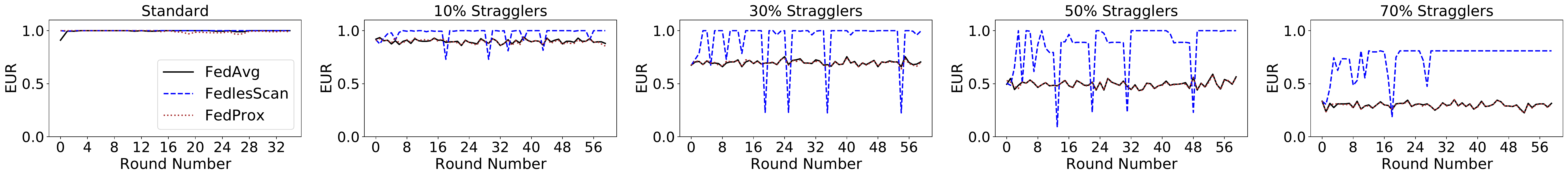}
\caption{Comparing \texttt{EUR}.}
\label{fig:eur-speech}
\end{subfigure}%
 \hfill
\begin{subfigure}{\textwidth}
\centering
\includegraphics[width=\textwidth]{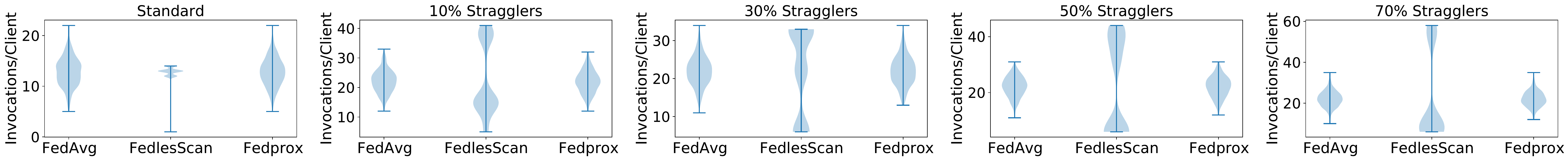}
\caption{Comparing the distribution of invocation frequency of clients.}
\label{fig:var-speech}
\end{subfigure}%
\caption{Comparing the different metrics for the different strategies on the Google Speech dataset~\cite{speech} with varying straggler ratios.}
\label{fig:metricsovertime}
\shrinkspace
\end{figure*}

\subsubsection{Experiment scenarios}
\label{sec:scenarios}

With our experiments, we aim to evaluate the performance of our strategy against delays and client function dropouts. Although the deployed functions will show delays, failures, and cold starts, it does not indicate how our strategy behaves in extreme situations, i.e., with a significantly high number of stragglers. Towards this, we consider two scenarios in our experiments.

\textbf{Standard Scenario}. In this, we perform the experiments on the deployed client functions without any modifications. Furthermore, we adjust the FL training round time to ensure that the clients can finish their training before the training round ends. This scenario demonstrates the performance in a more real-world situation when clients on commercial FaaS platforms are used for training.


\textbf{Straggler (\%) Scenario}. In this, we simulate varying percentages of stragglers in the FL system.  Although there might be different reasons for FaaS client failures, such as memory limit, function timeout, or communication issues, these failures can only have one of two effects on the clients. Clients can either push their updates after the training round is finished (slow updates) or can completely crash (not push their updates). To simulate slow updates, we limit the training round time to only fit clients with no issues or delays. Meaning, that clients which experience cold starts, bandwidth limitations, or communication delays do not finish the round in time; therefore, pushing their updates later. On the other hand, to simulate failures, we randomly select a specific ratio of clients to fail their training at the beginning of each experiment. We perform four different experiments for each dataset with 10\%, 30\%, 50\%, and 70\% stragglers in the system. Note that, in all our experiments, stragglers are different from \textit{malicious clients} found in some FL settings~\cite{li2020learning, zhang2022fldetector}. While malicious clients can act as stragglers to deliberately slow training or hinder model performance, they can perform more powerful attacks such as data or model poisoning~\cite{fl_challenges,fl_data_poisoning} to skew the model's performance. In our work, we focus on the problem of stragglers and their behavior rather than the clients' intentions. The number of training rounds for the different datasets and experiment scenarios is shown in Table~\ref{tab:experiment-params}.

\begin{table}
\centering

\begin{adjustbox}{width=8.8cm,  center}
\begin{tabular}{|c|c|c|c|c|c|c|c|c|c|c|c|} 
\hline
\multirow{3}{*}{\textbf{ Dataset}} & \multirow{3}{*}{\textbf{ Strategy}} & \multicolumn{10}{c|}{\textbf{Ratio of Stragglers}}                                                                                                                                          \\ 
\cline{3-12}
                                   &                                     & \multicolumn{2}{c|}{\textbf{Baseline}} & \multicolumn{2}{c|}{\textbf{10\%}} & \multicolumn{2}{c|}{\textbf{30\%}} & \multicolumn{2}{c|}{\textbf{50\%}} & \multicolumn{2}{c|}{\textbf{70\%}}  \\ 
\cline{3-12}
                                   &                                     & \textbf{Acc}   & \textbf{EUR}          & \textbf{Acc}   & \textbf{EUR}      & \textbf{Acc}   & \textbf{EUR}      & \textbf{Acc}   & \textbf{EUR}      & \textbf{Acc}   & \textbf{EUR}       \\ 
\hline
\multirow{3}{*}{MNIST}             & FedAvg                              & 0.981          & 0.99                  & \textbf{0.983} & 0.89              & \textbf{0.972} & 0.70               & 0.971          & 0.49              & 0.977          & 0.31               \\ 
\cline{2-12}
                                   & FedProx                             & 0.935          & 0.98                  & 0.972          & 0.88              & 0.960           & 0.69              & 0.962          & 0.49              & 0.967          & 0.29               \\ 
\cline{2-12}
                                   & FedLesScan                          & \textbf{0.985} & \textbf{0.99}         & 0.976          & \textbf{0.98}     & \textbf{0.972} & \textbf{0.96}     & \textbf{0.974} & \textbf{0.74}     & \textbf{0.979}  & \textbf{0.44}      \\ 
\hline
\multirow{3}{*}{FEMNIST}           & FedAvg                              & 0.756          & 0.99                  & 0.765          & 0.89              & 0.790           & 0.69              & 0.779         & 0.49              & 0.744           & 0.3                \\ 
\cline{2-12}
                                   & FedProx                             & 0.753          & 0.96                  & \textbf{0.778} & 0.80              & 0.756          & 0.64              & \textbf{0.785}  & 0.46              & 0.731           & 0.28               \\ 
\cline{2-12}
                                   & FedLesScan                          & \textbf{0.776} & \textbf{0.99}         & 0.770          & \textbf{0.97}     & \textbf{0.785} & \textbf{0.93}     & 0.774          & \textbf{0.80}      & \textbf{0.753} & \textbf{0.50}       \\ 
\hline
\multirow{3}{*}{Shakespeare}       & FedAvg                              & \textbf{0.434} & 0.87                  & \textbf{0.430}  & 0.80              & \textbf{0.401}   & 0.60               & 0.402   & 0.41              & 0.428           & 0.31               \\ 
\cline{2-12}
                                   & FedProx                             & 0.400          & 0.86                  & 0.399          & 0.78              & 0.396          & 0.58              & 0.345          & 0.40               & 0.38           & 0.29               \\ 
\cline{2-12}
                                   & FedLesScan                          & 0.388          & \textbf{0.94}         & 0.408          & \textbf{0.90}      & 0.400          & \textbf{0.86}     & \textbf{0.403}   & \textbf{0.72}     & \textbf{0.432}  & \textbf{0.53}      \\ 
\hline
\multirow{3}{*}{Google Speech}     & FedAvg                              & 0.766           & \textbf{0.99}        & 0.699            & 0.90               & 0.639           & 0.70               & 0.645           & 0.50               & 0.804           & 0.30                \\ 
\cline{2-12}
                                   & FedProx                             & 0.774          & \textbf{0.99}         & 0.664           & 0.89              & 0.709           & 0.70               & 0.694           & 0.49              & 0.759           & 0.29               \\ 
\cline{2-12}
                                   & FedLesScan                          & \textbf{0.794} & \textbf{0.99}         & \textbf{0.762}   & \textbf{0.97}     & \textbf{0.719}  & \textbf{0.90}      & \textbf{0.720}  & \textbf{0.86}     & \textbf{0.824}  & \textbf{0.74}      \\
\hline
\end{tabular}
\end{adjustbox}
\caption{Accuracy and EUR values for the three strategies across different scenarios (\S\ref{sec:scenarios}) and datasets (\S\ref{sec:datasets}). The highlighted values represent the highest accuracy and EUR values for a particular dataset, strategy and scenario.}
\label{tab:acc-all}
\shrinkspace
\vspace{-1mm}
\end{table}

\subsubsection{Metrics}
\label{sec:metrics}
For comparing the different strategies, we use the standard metrics \textit{accuracy}, \textit{experiment duration}, and \textit{cost}. For calculating accuracy of the trained global model, we randomly choose a set of clients and evaluate it on their test datasets. Following this, we multiply the obtained accuracy for a particular client by the ratio of its test set cardinality to the total cardinality of the test dataset. The final accuracy value is obtained by averaging the obtained accuracy values. While accuracy describes model performance, it does not provide insights about the performance of the strategy and the contributions of the clients to the global model. Towards this, we also use the metrics \textit{Effective Update Ratio} (EUR)~\cite{safa} and \textit{Bias}~\cite{safa}. EUR is defined as the ratio between the successful clients and the subset of selected clients. It shows the effect of stragglers on round utilization. A higher value of EUR represents less wasted resources since clients requested to participate in a certain round end up contributing to the global model. Furthermore, to provide insights into the bias of the client selection schemes, we use variance plots. This is done by showing the frequency of selection for each client across the FL training process. Bias is defined as the difference between the frequency of the least called client and the most called client~\cite{safa}. For scenarios with low number of stragglers, we target low bias, while for scenarios with high number of stragglers the bias should be higher due to the prioritization of reliable clients in training. Experiment duration represents the total time required for training the model. For computing training costs, we use the cost computation model~\cite{gcf_cost} used by Google to estimate the cost for each client function based on the number of invocations, allocated memory, and execution duration.


\subsection{Comparing accuracy and round utilization}
\label{sec:accandutil}

 
The obtained accuracy and EUR values across all our experiments is summarized in Table~\ref{tab:acc-all}. For the \textit{standard scenarios} (\S\ref{sec:scenarios}), we obtained better accuracy and round utlization for our strategy as compared to \texttt{FedAvg} and \texttt{FedProx} across all datasets except Shakespeare. This is because with the Shakespeare dataset, some of the clients contribute to the model accuracy more than others, especially clients with longer training times. Moreover, due to budget constraints and significantly high training costs, we train for a slightly less number of rounds, i.e., 25 for Shakespeare. However, we argue that in a more realistic scenario with more number of rounds, the difference in accuracy will decrease. This is because our strategy utilizes clients more efficiently, i.e., higher EUR. As a result, with more rounds the number of invocations per client will increase and more clients will contribute to the global model, leading to a higher accuracy. Similarly, for the \textit{straggler (\%) scenarios}, we obtained better results for accuracy with our strategy as compared to the other two for the MNIST, FEMNIST, and the Google Speech datasets. On the Shakespeare dataset, our strategy outperformed \texttt{FedAvg} and \texttt{FedProx} in scenarios with 30\%, 50\%, and 70\% stragglers. In terms of EUR, our strategy constantly outperforms the other two strategies across all scenarios and datasets since they use random selection for selecting clients for an FL training round.

For most \textit{standard scenarios} (Table~\ref{tab:acc-all}), \texttt{FedLesScan} obtained better accuracy as compared to the other two strategies due to the better distribution of client invocations. This is because our strategy prioritizes clients with the least number of invocations in a selected client cluster (\S\ref{sec:ClientSelection}) leading to more balanced contributions among the participating clients. On the other hand, with \textit{straggler (\%) scenario}, our strategy reached better accuracies by relying more on robust and reliable clients. Furthermore, the utilization of a staleness-aware aggregation scheme (\S\ref{sec:stalelessaware}) avoids wasting valuable contributions, which in turn increases accuracy.

\begin{table}[t]
\centering
\begin{adjustbox}{width=8.5cm,  center}
\begin{tabular}{|c|c|c|c|c|c|c|} 
\hline
\multirow{2}{*}{\textbf{ Dataset}} & \multirow{2}{*}{\textbf{ Strategy}} & \multicolumn{5}{c|}{\textbf{Experiment Time (mins)}}                               \\ 
\cline{3-7}
                                   &                                     & \textbf{Standard} & \textbf{10\%} & \textbf{30\%} & \textbf{50\%} & \textbf{70\%}  \\ 
\hline
\multirow{3}{*}{MNIST}             & FedAvg                              & 39.7              & 40.2          & 40.0            & \textbf{40.0}            & \textbf{40.0}             \\ 
\cline{2-7}
                                   & FedProx                             & 40.0                & 40.3          & 40.0          & \textbf{40.0}            & \textbf{40.0}             \\ 
\cline{2-7}
                                   & FedLesScan                          & \textbf{23.7}              & \textbf{28.6}          & \textbf{27.3}          & \textbf{40.0}            & \textbf{40.0}             \\ 
\hline
\multirow{3}{*}{FEMNIST}           & FedAvg                              & 75.5              & 86.7          & 86.9          & \textbf{86.7}          & \textbf{86.7}          \\ 
\cline{2-7}
                                   & FedProx                             & 112.4               & 88.2            & 88.1            & 87.8            & 87.4             \\ 
\cline{2-7}
                                   & FedLesScan                          & \textbf{70.9}              & \textbf{75.6}          & \textbf{82.8}          & 86.8          & \textbf{86.7}           \\ 
\hline
\multirow{3}{*}{Shakespeare}       & FedAvg                              & 217.0           & 217.0        & 217.0         & 216.9         & 216.8          \\ 
\cline{2-7}
                                   & FedProx                             & 217.0             & 217.0         & 217.0        & 217.0        & 216.7          \\ 
\cline{2-7}
                                   & FedLesScan                          & \textbf{185.5}             &\textbf{ 215.5  }         & \textbf{205.2}           & \textbf{216.8}           & \textbf{216.6}            \\ 
\hline
\multirow{3}{*}{Google Speech}    & FedAvg                              & 20.3                & 40.1            & 40.1            & 40.0           & \textbf{40.0}             \\ 
\cline{2-7}
                                   & FedProx                             & 21.5              & 40.0           & 40.0           & 40.0           & \textbf{40.0}             \\ 
\cline{2-7}
                                   & FedLesScan                          & \textbf{15.1}                & \textbf{31.1}            & \textbf{28.8}            & \textbf{33.3}            & \textbf{40.0}             \\
\hline
\end{tabular}
\end{adjustbox}
\caption{Comparing total time for the three strategies across different scenarios (\S\ref{sec:scenarios}) and datasets (\S\ref{sec:datasets}). The highlighted values represent the minimum training time for each experiment.}
\label{tab:total-time}
\shrinkspace

\end{table}

To offer detailed insights, we present results for the Google Speech dataset~\cite{speech} wrt the metrics accuracy and EUR, across the FL training session as shown in Figures \ref{fig:acc-speech} and \ref{fig:eur-speech} respectively. For the \textit{standard scenarios}, we ran the FL training session for $35$ rounds, while for the \textit{straggler (\%) scenarios}, we ran the experiments for $60$ rounds. For the standard scenario, our strategy reached an accuracy of 79.4\% as compared to 76.6\% and 77.4\% for \texttt{FedAvg} and \texttt{FedProx} respectively. Furthermore, our strategy showed faster convergence by reaching an accuracy of 70\% in 19 rounds as compared to 21 and 22 for \texttt{FedAvg} and \texttt{FedProx} respectively. With 10\% stragglers, our strategy and \texttt{FedAvg} had a similar convergence rate, while \texttt{FedProx} was slightly behind. Moreover, our strategy reached an end accuracy of 76\% which is a 6\% and 10\% increase over \texttt{FedAvg} and \texttt{FedProx} respectively. For an FL system with 30\% stragglers, our strategy consistently outperforms \texttt{FedAvg} by around 8\% towards the end of the training, while outperforming \texttt{FedProx} by a smaller margin of 1\%. We observe a similar trend for our experiments with 50\% and 70\% stragglers in the system, where our strategy outperforms the other two. From our experiments, we observe that the presence of stragglers affects the convergence speed of the DNN models in an FL training session. DNN models trained in \textit{standard scenarios} converge faster as compared to the models trained with \textit{straggler (\%) scenarios} as shown in Figure~\ref{fig:acc-speech} (Table~\ref{tab:experiment-params}). We observe a similar behaviour for the different datasets (\S\ref{sec:datasets}). Moreover, for the \textit{straggler (\%) scenarios}, increasing the percentage of stragglers in the system does not consistently decrease the accuracy of the trained DNN model as shown in Table~\ref{tab:acc-all}. This is especially true for experiments with a higher number of stragglers, i.e., $70\%$. This behavior is not exclusive to our experiments and was also reported by Li et al.~\cite{fedprox} and Wu et al.~\cite{safa}. While the authors do not provide a clear explanation for this behaviour, we argue that due to the non-IID nature of the data, clients do not contribute evenly to the test accuracy. In addition, having fewer number of reliable clients reduces the varying effect of local model updates and local model deviations from the global model, making it easier to reach a consensus on the global model. Therefore, we can reach situations where a system with more stragglers reaches higher overall accuracy at the end.

\begin{table}[t]
\centering
\begin{adjustbox}{width=8.5cm,  center}
\begin{tabular}{|c|c|c|c|c|c|c|} 
\hline
\multirow{2}{*}{\textbf{ Dataset}} & \multirow{2}{*}{\textbf{ Strategy}} & \multicolumn{5}{c|}{\textbf{Experiment Cost (\$)}}                                 \\ 
\cline{3-7}
                                   &                                     & \textbf{Standard} & \textbf{10\%} & \textbf{30\%} & \textbf{50\%} & \textbf{70\%}  \\ 
\hline
\multirow{3}{*}{MNIST}             & FedAvg                              & 2.90               & 3.90           & 6.00             & 8.00             & 10.40           \\ 
\cline{2-7}
                                   & FedProx                             & 5.50               & 6.40           & 7.60           & 9.21          & 11.03          \\ 
\cline{2-7}
                                   & FedLesScan                          & \textbf{2.70}               & \textbf{3.86}           & \textbf{4.00}             & \textbf{5.99}             &\textbf{9.2}           \\ 
\hline
\multirow{3}{*}{FEMNIST}           & FedAvg                              & 13.50              & 16.19          & 17.87         & 20.54         & 24.70           \\ 
\cline{2-7}
                                   & FedProx                             & 16.67              & 17.29          & 19.40          & 22.42         & 25.80           \\ 
\cline{2-7}
                                   & FedLesScan                          & \textbf{13.17}              & \textbf{14.58}         & \textbf{14.40}          & \textbf{14.81}         & \textbf{20.60}           \\ 
\hline
\multirow{3}{*}{Shakespeare}       & FedAvg                              & 5.40               & 6.60           & 9.21           & 12.50          & 15.40           \\ 
\cline{2-7}
                                   & FedProx                             & \textbf{5.12}              & 6.72          & 9.00             & 12.20          & 15.40           \\ 
\cline{2-7}
                                   & FedLesScan                          & 5.33              & \textbf{5.50}           & \textbf{6.75}          &\textbf{ 8.46}          & \textbf{12.00}             \\ 
\hline
\multirow{3}{*}{Google Speech}    & FedAvg                              & 1.98              & 3.90         & 6.40         & 8.30         & 10.50           \\ 
\cline{2-7}
                                   & FedProx                             & 2.39             & 4.60         & 6.77          & 8.70          & 10.80           \\ 
\cline{2-7}
                                   & FedLesScan                          & \textbf{1.73}              & \textbf{2.70}         & \textbf{3.68}          & \textbf{4.20}          & \textbf{5.50}           \\
\hline
\end{tabular}
\end{adjustbox}
\caption{Comparing training costs for the three strategies across different scenarios (\S\ref{sec:scenarios}) and datasets (\S\ref{sec:datasets}). The highlighted values represent the minimum experiment cost.}
\label{tab:cost-analysis}
\shrinkspace
\end{table}

 Figure~\ref{fig:eur-speech} shows the EUR comparison among the three strategies for the different scenarios. In the standard scenario, the three strategies performed similarly, achieving more than 99\% average EUR ratio during training. In the straggler (\%) scenarios, our strategy consistently achieves higher EUR as compared to the other strategies. Furthermore, as the number of stragglers in the system increases, the difference between the average EUR of our strategy as compared to the other two also increases. We observe occasional drops in EUR for our strategy with varying number of stragglers in the system. These drops demonstrate the effect of clustering of the clients. Distributing slow clients across the training rounds will affect the efficiency of more rounds. 
 

Our strategy utilizes dynamic clusters to combine clients with similar behavior together. Recomputing the clusters each round based on clients'  recent behavior reduces the impact of stragglers on training the other clients. However, this leads to occasional drops in the EUR, which happens when an unreliable subset of clients is invoked. The subsequent rounds, involving the rest of the clients, maintain higher EUR values, thereby decreasing the total training time.

Although EUR demonstrates the efficiency of the system, it does not show the bias of the strategy. A system that utilizes a specific subset of clients and discards the rest will have a higher EUR but will underutilize the rest of the clients. To this end, we use violin plots as shown in Figure~\ref{fig:var-speech} to provide insights into the bias encountered by our strategy (\S\ref{sec:metrics}). The graph shows a distribution based on the number of invocations for each client (y-axis). We demonstrate bias by the difference between the highest and lowest points in the distribution. A greater difference (height) represents that the algorithm is biased towards a specific subset of clients, while a smaller difference (height) represents that the difference between the most and least invoked clients is low. A bigger width indicates that certain clients were invoked more frequently. For the \textit{standard scenario}, we observe that \texttt{FedAvg} and \texttt{FedProx} show similar behaviour without distinguishing between stragglers and reliable clients. This is because they use random client selection. On the other hand, our strategy adapts to stragglers and promotes fair selection of clients. This is apparent since the distribution of our strategy is centered around similar values, i.e., most clients have the same number of invocations. Furthermore, we observe that with our strategy few clients have a low number of invocations. This is because clients that have repeatedly failed due to memory constraints are not used as often as the rest of the clients. For the \textit{straggler (\%) scenarios}, we observe that our strategy prioritizes reliable clients while relying on stragglers less during training. We see a distinction between the number of invocations of reliable clients and stragglers.




From our experiments, we observe that \texttt{FedLesScan} performs best in long FL training sessions due to the large amount of training data collected about the behaviour of the clients. The collected data leads to better groupings of the clients (\S\ref{sec:ClientSelection}). Moreover, in the case of repeated failures of clients, our strategy should perform well due to its three-tier classification system (\S\ref{sec:tiers}). 


\subsection{Comparing time and cost}
\label{sec:timecost}
Although model accuracy is an important metric, fast convergence wrt the number of training rounds does not provide a complete picture of the efficiency of the system. Towards this, in this section, we provide a collective analysis of all experiments wrt total duration and costs. To compute the total experiment duration, we aggregate the total round time during training. For all the three strategies, the round time is determined by the slowest invoked client. As a result, the round time depends on either the response of the slowest client or a predetermined timeout (\S\ref{sec:scenarios}). Furthermore, for \textit{straggler (\%) scenarios}, we simulate real-world behavior by assuming that the stragglers will not respond, thus forcing the controller (\S\ref{sec:sysdesign}) to wait until the round timeout. 




Table~\ref{tab:total-time} shows the total aggregated time per experiment. For the \textit{standard scenario}, training a model with our strategy is significantly faster as compared to \texttt{FedAvg} and \texttt{FedProx} across all datasets. For instance, for the MNIST dataset, our strategy takes 40\% less time as compared to the other strategies. For the \textit{straggler (\%) scenarios}, we see the effect of stragglers on experiment duration. In the scenarios with 10\% and 30\% stragglers, we observe that 
\texttt{FedLesScan} maintains a lower duration across all experiments. However, when the number of stragglers in the system is significantly higher, they must be invoked in almost all training rounds to meet the minimum number of clients required per round. We notice this behavior for our strategy with greater than 50\% stragglers in the system for the MNIST, FEMNIST, and the Shakespeare dataset. However, for the Google Speech dataset, our strategy still has an 18\% lower experiment duration as compared to the other two. This is because the total number of clients for the Google Speech dataset is 542 with 200 concurrent clients participating in a training round (\S\ref{sec:expconfig}). For the scenario with 70\% stragglers in the system, all approaches have similar experiment times across all datasets.

To analyze the cost of the experiments, we had to estimate the cost of stragglers since stragglers can either miss their round or fail. However, their running cost still factors into the experiment cost. In the worst-case scenario, stragglers can increase costs by wasting resources doing computations on wasted contributions. As a result,  we estimate the cost of stragglers as the cost of running the functions for the entire round duration. As described in \S\ref{sec:metrics}, we use the cost model for Google Cloud Functions~\cite{gcf_cost} for calculating the cost of the clients. 

Table~\ref{tab:cost-analysis} shows the cost for the different strategies, datasets, and scenarios. For the \textit{standard scenario} with MNIST, we observe a 6.8\% and 50\% cost reduction for our strategy as compared to \texttt{FedAvg} and \texttt{FedProx}. Similarly for the FEMNIST and Google Speech datasets in the \textit{standard scenario}, we observe cost reductions of about 2\%, 20\% and 12\%, 27\% as compared to \texttt{FedAvg} and \texttt{FedProx} respectively. On the other hand, for the Shakespeare dataset, \texttt{FedProx} performed better than our strategy and \texttt{FedAvg} by 4\% and 6\% respectively. We observe that the efficiency of our strategy becomes more visible as the number of stragglers in the system increases. For all the scenarios with a varying number of stragglers, our strategy has the minimum cost as compared to the other two strategies. For the \textit{straggler (\%) scenarios}, our strategy consistently achieved lower experiment cost across all datasets with an average cost reduction of 25\% and 32\% as compared to \texttt{FedAvg} and \texttt{FedProx} respectively.

\section{Conclusion \& Future Work}
\label{sec:conclusion}
In this paper, we made two main contributions. First, we made several extensions to an open-source system and framework for serverless FL called \texttt{FedLess}. Towards this, we made it easier to use and deploy, modified its architectural components that enabled running multiple training strategies, and integrated a mocking system that can simulate the behavior of all system components for easier development and debugging. To this end, \texttt{FedLess} represents a real system that can be used by data holders for distributed training of models via serverless FL. Second, we proposed and implemented \texttt{FedLesScan}, a novel clustering-based training strategy designed for FL on FaaS platforms. It includes an intelligent client selection algorithm based on clustering clients with similar behavior together. Our strategy dynamically adapts to the clients' behavior and provides better overall system utilization. Furthermore, we integrated a staleness-aware aggregation mechanism to mitigate wasted contributions for slow clients. We comprehensively evaluated \texttt{FedLesScan} by comparing its performance against two popular strategies on multiple datasets. Furthermore, we analyzed its behavior in different system states with varying ratio of stragglers. Overall, our experiments showed that \texttt{FedLesScan} achieved  better results in terms of accuracy, training time, effective update ratio, and training costs by better utilizing the participating clients. In the future, we plan to investigate the use of more advanced staleness-aware aggregation schemes that aggregate valuable updates and discard the unnecessary ones. Moreover, we plan to explore dynamically adapting the number of clients selected each round in \texttt{FedLesScan} based on the current system state.

\section{Acknowledgement}
The research leading to these results was funded by the Deutsche Forschungsgemeinschaft (DFG, German Research Foundation)-Proje\-ktnummer 146371743-TRR 89: Invasive Computing. Moreover, this work was supported in partial by the funding of the German Federal Ministry of Education and Research (BMBF) in the scope of the Software Campus program. Google Cloud credits in this work were provided by the \textit{Google Cloud Research Credits} program with the award number 64c92de5-fb62-4386-8c5b-ff3f480390bb. All code artifacts related to this work can be found here\footnote{\url{https://tinyurl.com/3utdhuuu}}.

\bibliographystyle{IEEEtran}
\thispagestyle{empty}
\bibliography{parallelpgm}

\begin{thebibliography}{10}
\providecommand{\url}[1]{#1}
\csname url@samestyle\endcsname
\providecommand{\newblock}{\relax}
\providecommand{\bibinfo}[2]{#2}
\providecommand{\BIBentrySTDinterwordspacing}{\spaceskip=0pt\relax}
\providecommand{\BIBentryALTinterwordstretchfactor}{4}
\providecommand{\BIBentryALTinterwordspacing}{\spaceskip=\fontdimen2\font plus
\BIBentryALTinterwordstretchfactor\fontdimen3\font minus
  \fontdimen4\font\relax}
\providecommand{\BIBforeignlanguage}[2]{{%
\expandafter\ifx\csname l@#1\endcsname\relax
\typeout{** WARNING: IEEEtran.bst: No hyphenation pattern has been}%
\typeout{** loaded for the language `#1'. Using the pattern for}%
\typeout{** the default language instead.}%
\else
\language=\csname l@#1\endcsname
\fi
#2}}
\providecommand{\BIBdecl}{\relax}
\BIBdecl

\bibitem{chiang2016fog}
M.~Chiang and T.~Zhang, ``Fog and iot: An overview of research opportunities,''
  \emph{IEEE Internet of things journal}, vol.~3, no.~6, pp. 854--864, 2016.

\bibitem{NvidiaClara2020}
\BIBentryALTinterwordspacing
{Nvidia Clara}, ``{NVIDIA Clara | NVIDIA Developer},'' 2020. [Online].
  Available: \url{https://developer.nvidia.com/blog/federated-learning-clara/}
\BIBentrySTDinterwordspacing

\bibitem{resnet}
K.~He, X.~Zhang, S.~Ren, and J.~Sun, ``Deep residual learning for image
  recognition,'' in \emph{2016 IEEE Conference on Computer Vision and Pattern
  Recognition (CVPR)}, 2016, pp. 770--778.

\bibitem{EUdataregulations2018}
\BIBentryALTinterwordspacing
{2018 reform of EU data protection rules}. [Online]. Available:
  \url{https://eur-lex.europa.eu/eli/reg/2016/679/oj}
\BIBentrySTDinterwordspacing

\bibitem{mcmahan2017communication}
B.~McMahan, E.~Moore, D.~Ramage, S.~Hampson, and B.~A. y~Arcas,
  ``Communication-efficient learning of deep networks from decentralized
  data,'' in \emph{Artificial Intelligence and Statistics}.\hskip 1em plus
  0.5em minus 0.4em\relax PMLR, 2017, pp. 1273--1282.

\bibitem{lecun2015deep}
Y.~LeCun, Y.~Bengio, and G.~Hinton, ``{Deep learning},'' \emph{nature}, vol.
  521, no. 7553, pp. 436--444, 2015.

\bibitem{paddle}
\BIBentryALTinterwordspacing
``{PaddlePaddle/PaddleFL: Federated Deep Learning in PaddlePaddle}.'' [Online].
  Available: \url{https://github.com/PaddlePaddle/PaddleFL}
\BIBentrySTDinterwordspacing

\bibitem{flower}
\BIBentryALTinterwordspacing
D.~J. Beutel, T.~Topal, A.~Mathur, X.~Qiu, T.~Parcollet, N.~D. Lane, P.~P.~B.
  de~Gusm{\~{a}}o, and N.~D. Lane, ``{Flower: A Friendly Federated Learning
  Research Framework},'' \emph{arXiv}, pp. 1--22, jul 2020. [Online].
  Available: \url{http://arxiv.org/abs/2007.14390}
\BIBentrySTDinterwordspacing

\bibitem{fedml}
\BIBentryALTinterwordspacing
``{FedML}.'' [Online]. Available: \url{https://github.com/FedML-AI/FedML}
\BIBentrySTDinterwordspacing

\bibitem{serverlessfl}
\BIBentryALTinterwordspacing
M.~Chadha, A.~Jindal, and M.~Gerndt, ``Towards federated learning using faas
  fabric,'' in \emph{Proceedings of the 2020 Sixth International Workshop on
  Serverless Computing}, ser. WoSC'20.\hskip 1em plus 0.5em minus 0.4em\relax
  New York, NY, USA: Association for Computing Machinery, 2020, p. 49–54.
  [Online]. Available: \url{https://doi.org/10.1145/3429880.3430100}
\BIBentrySTDinterwordspacing

\bibitem{fedless}
\BIBentryALTinterwordspacing
A.~Grafberger, M.~Chadha, A.~Jindal, J.~Gu, and M.~Gerndt, ``Fedless: Secure
  and scalable federated learning using serverless computing,'' in \emph{2021
  IEEE International Conference on Big Data (Big Data)}, 2021, pp. 164--173.
  [Online]. Available: \url{https://doi.org/10.1109/BigData52589.2021.9672067}
\BIBentrySTDinterwordspacing

\bibitem{jayaram2022lambda}
\BIBentryALTinterwordspacing
K.~Jayaram, V.~Muthusamy, G.~Thomas, A.~Verma, and M.~Purcell, ``$\lambda$-fl:
  Serverless aggregation for federated learning,'' 2022. [Online]. Available:
  \url{https://federated-learning.org/fl-aaai-2022/Papers/FL-AAAI-22_paper_44.pdf}
\BIBentrySTDinterwordspacing

\bibitem{cncf-serverless-whitepaper}
\BIBentryALTinterwordspacing
S.~Allen, B.~Browning, L.~Calcote, A.~Chaudhry, D.~Davis, L.~Fourie, A.~Gulli,
  Y.~Haviv, D.~Krook, O.~Nissan-Messing, C.~Munns, K.~Owens, M.~Peek, C.~Zhang,
  and C.~A., ``{CNCF WG-Serverless Whitepaper v1.0},'' CNCF, Tech. Rep., 2018.
  [Online]. Available:
  \url{https://github.com/cncf/wg-serverless/blob/master/whitepapers/serverless-overview/cncf_serverless_whitepaper_v1.0.pdf}
\BIBentrySTDinterwordspacing

\bibitem{gcloud-functions-2}
\BIBentryALTinterwordspacing
{Google Cloud}, ``{Cloud Functions Second Generation| Google Cloud},'' 2022.
  [Online]. Available:
  \url{https://cloud.google.com/functions/docs/2nd-gen/overview}
\BIBentrySTDinterwordspacing

\bibitem{bonawitz2019federated}
K.~Bonawitz, H.~Eichner, W.~Grieskamp, D.~Huba, A.~Ingerman, V.~Ivanov,
  C.~Kiddon, J.~Konečný, S.~Mazzocchi, H.~B. McMahan, T.~V. Overveldt,
  D.~Petrou, D.~Ramage, and J.~Roselander, ``Towards federated learning at
  scale: System design,'' 2019.

\bibitem{jayaram2022adaptive}
\BIBentryALTinterwordspacing
K.~Jayaram, V.~Muthusamy, G.~Thomas, A.~Verma, and M.~Purcell, ``Adaptive
  aggregation for federated learning,'' \emph{arXiv preprint arXiv:2203.12163},
  2022. [Online]. Available: \url{https://arxiv.org/pdf/2203.12163.pdf}
\BIBentrySTDinterwordspacing

\bibitem{jayaram2022just}
\BIBentryALTinterwordspacing
K.~Jayaram, A.~Verma, G.~Thomas, and V.~Muthusamy, ``Just-in-time aggregation
  for federated learning,'' \emph{arXiv preprint arXiv:2208.09740}, 2022.
  [Online]. Available: \url{https://arxiv.org/pdf/2208.09740.pdf}
\BIBentrySTDinterwordspacing

\bibitem{speech}
\BIBentryALTinterwordspacing
P.~Warden, ``Speech commands: A dataset for limited-vocabulary speech
  recognition,'' 2018. [Online]. Available:
  \url{https://arxiv.org/abs/1804.03209}
\BIBentrySTDinterwordspacing

\bibitem{flmec}
W.~Y.~B. Lim, N.~C. Luong, D.~T. Hoang, Y.~Jiao, Y.-C. Liang, Q.~Yang,
  D.~Niyato, and C.~Miao, ``Federated learning in mobile edge networks: A
  comprehensive survey,'' \emph{IEEE Communications Surveys Tutorials},
  vol.~22, no.~3, pp. 2031--2063, 2020.

\bibitem{fedprox}
A.~K. Sahu, T.~Li, M.~Sanjabi, M.~Zaheer, A.~Talwalkar, and V.~Smith, ``On the
  convergence of federated optimization in heterogeneous networks,''
  \emph{arXiv preprint arXiv:1812.06127}, vol.~3, p.~3, 2018.

\bibitem{gradientreduced}
\BIBentryALTinterwordspacing
C.~T. Dinh, N.~H. Tran, T.~D. Nguyen, W.~Bao, A.~Y. Zomaya, and B.~B. Zhou,
  ``Federated learning with proximal stochastic variance reduced gradient
  algorithms,'' in \emph{49th International Conference on Parallel Processing -
  ICPP}, ser. ICPP '20.\hskip 1em plus 0.5em minus 0.4em\relax New York, NY,
  USA: Association for Computing Machinery, 2020. [Online]. Available:
  \url{https://doi.org/10.1145/3404397.3404457}
\BIBentrySTDinterwordspacing

\bibitem{fed_async}
C.~Xie, S.~Koyejo, and I.~Gupta, ``Asynchronous federated optimization,''
  \emph{arXiv preprint arXiv:1903.03934}, 2019.

\bibitem{asofed}
\BIBentryALTinterwordspacing
Y.~Chen, Y.~Ning, M.~Slawski, and H.~Rangwala, ``Asynchronous online federated
  learning for edge devices with non-iid data,'' in \emph{2020 IEEE
  International Conference on Big Data (Big Data)}.\hskip 1em plus 0.5em minus
  0.4em\relax Los Alamitos, CA, USA: IEEE Computer Society, dec 2020, pp.
  15--24. [Online]. Available:
  \url{https://doi.ieeecomputersociety.org/10.1109/BigData50022.2020.9378161}
\BIBentrySTDinterwordspacing

\bibitem{fedAt}
\BIBentryALTinterwordspacing
Z.~Chai, Y.~Chen, A.~Anwar, L.~Zhao, Y.~Cheng, and H.~Rangwala, ``Fedat: A
  high-performance and communication-efficient federated learning system with
  asynchronous tiers,'' in \emph{Proceedings of the International Conference
  for High Performance Computing, Networking, Storage and Analysis}, ser. SC
  '21.\hskip 1em plus 0.5em minus 0.4em\relax New York, NY, USA: Association
  for Computing Machinery, 2021. [Online]. Available:
  \url{https://doi.org/10.1145/3458817.3476211}
\BIBentrySTDinterwordspacing

\bibitem{csafl}
Y.~Zhang, M.~Duan, D.~Liu, L.~Li, A.~Ren, X.~Chen, Y.~Tan, and C.~Wang,
  ``Csafl: A clustered semi-asynchronous federated learning framework,'' in
  \emph{2021 International Joint Conference on Neural Networks (IJCNN)}.\hskip
  1em plus 0.5em minus 0.4em\relax IEEE, 2021, pp. 1--10.

\bibitem{safa}
W.~Wu, L.~He, W.~Lin, R.~Mao, C.~Maple, and S.~Jarvis, ``Safa: a
  semi-asynchronous protocol for fast federated learning with low overhead,''
  \emph{IEEE Transactions on Computers}, vol.~70, no.~5, pp. 655--668, 2020.

\bibitem{fleet}
G.~Damaskinos, R.~Guerraoui, A.-M. Kermarrec, V.~Nitu, R.~Patra, and F.~Taiani,
  ``Fleet: Online federated learning via staleness awareness and performance
  prediction,'' in \emph{Proceedings of the 21st International Middleware
  Conference}, 2020, pp. 163--177.

\bibitem{fedskel}
\BIBentryALTinterwordspacing
J.~Luo, J.~Yang, X.~Ye, X.~Guo, and W.~Zhao, ``Fedskel: Efficient federated
  learning on heterogeneous systems with skeleton gradients update,'' in
  \emph{Proceedings of the 30th ACM International Conference on Information \&
  Knowledge Management}, ser. CIKM '21.\hskip 1em plus 0.5em minus 0.4em\relax
  New York, NY, USA: Association for Computing Machinery, 2021, p. 3283–3287.
  [Online]. Available: \url{https://doi.org/10.1145/3459637.3482107}
\BIBentrySTDinterwordspacing

\bibitem{wang2018peeking}
L.~Wang, M.~Li, Y.~Zhang, T.~Ristenpart, and M.~Swift, ``Peeking behind the
  curtains of serverless platforms,'' in \emph{2018 $\{$USENIX$\}$ Annual
  Technical Conference ($\{$USENIX$\}$$\{$ATC$\}$ 18)}, 2018, pp. 133--146.

\bibitem{hep}
\BIBentryALTinterwordspacing
J.~Kuundefinednierz, M.~Malawski, V.~E. Padulano, E.~Tejedor~Saavedra, and
  P.~Alonso-Jorda, ``Distributed parallel analysis engine for high energy
  physics using aws lambda,'' in \emph{Proceedings of the 1st Workshop on High
  Performance Serverless Computing}, ser. HiPS '21.\hskip 1em plus 0.5em minus
  0.4em\relax New York, NY, USA: Association for Computing Machinery, 2020, p.
  13–16. [Online]. Available: \url{https://doi.org/10.1145/3452413.3464788}
\BIBentrySTDinterwordspacing

\bibitem{chadha2021architecture}
\BIBentryALTinterwordspacing
M.~Chadha, A.~Jindal, and M.~Gerndt, ``Architecture-specific performance
  optimization of compute-intensive faas functions,'' in \emph{2021 IEEE 14th
  International Conference on Cloud Computing (CLOUD)}, 2021, pp. 478--483.
  [Online]. Available: \url{https://doi.org/10.1109/CLOUD53861.2021.00062}
\BIBentrySTDinterwordspacing

\bibitem{Carreira2019}
J.~Carreira, P.~Fonseca, A.~Tumanov, A.~Zhang, and R.~Katz, ``{Cirrus: A
  Serverless Framework for End-To-end ML Workflows},'' in \emph{SoCC 2019 -
  Proceedings of the ACM Symposium on Cloud Computing}.\hskip 1em plus 0.5em
  minus 0.4em\relax Association for Computing Machinery, nov 2019, pp. 13--24.

\bibitem{tinyfaas}
T.~Pfandzelter and D.~Bermbach, ``tinyfaas: A lightweight faas platform for
  edge environments,'' in \emph{2020 IEEE International Conference on Fog
  Computing (ICFC)}, 2020, pp. 17--24.

\bibitem{fado}
\BIBentryALTinterwordspacing
C.~P. Smith, A.~Jindal, M.~Chadha, M.~Gerndt, and S.~Benedict, ``Fado: Faas
  functions and data orchestrator for multiple serverless edge-cloud
  clusters,'' in \emph{2022 IEEE 6th International Conference on Fog and Edge
  Computing (ICFEC)}, 2022, pp. 17--25. [Online]. Available:
  \url{https://doi.org/10.1109/ICFEC54809.2022.00010}
\BIBentrySTDinterwordspacing

\bibitem{courier}
\BIBentryALTinterwordspacing
A.~Jindal, J.~Frielinghaus, M.~Chadha, and M.~Gerndt, ``Courier: Delivering
  serverless functions within heterogeneous faas deployments,'' in \emph{2021
  IEEE/ACM 14th International Conference on Utility and Cloud Computing
  (UCC'21)}, ser. UCC '21.\hskip 1em plus 0.5em minus 0.4em\relax New York, NY,
  USA: Association for Computing Machinery, 2021. [Online]. Available:
  \url{https://doi.org/10.1145/3468737.3494097}
\BIBentrySTDinterwordspacing

\bibitem{fncapacitor}
\BIBentryALTinterwordspacing
A.~Jindal, M.~Chadha, S.~Benedict, and M.~Gerndt, ``Estimating the capacities
  of function-as-a-service functions,'' in \emph{Proceedings of the 14th
  IEEE/ACM International Conference on Utility and Cloud Computing Companion},
  ser. UCC '21.\hskip 1em plus 0.5em minus 0.4em\relax New York, NY, USA:
  Association for Computing Machinery, 2022. [Online]. Available:
  \url{https://doi.org/10.1145/3492323.3495628}
\BIBentrySTDinterwordspacing

\bibitem{docker}
\BIBentryALTinterwordspacing
``{Docker containers},'' 2022. [Online]. Available:
  \url{https://www.docker.com/}
\BIBentrySTDinterwordspacing

\bibitem{aws-lambda}
\BIBentryALTinterwordspacing
Aws, ``{AWS Lambda – Serverless Compute - Amazon Web Services},'' 2020.
  [Online]. Available: \url{https://aws.amazon.com/lambda/}
\BIBentrySTDinterwordspacing

\bibitem{openfaas}
\BIBentryALTinterwordspacing
OpenFaaS, ``{OpenFaaS - Serverless Functions Made Simple},'' 2019. [Online].
  Available: \url{https://www.openfaas.com/ https://docs.openfaas.com/}
\BIBentrySTDinterwordspacing

\bibitem{agile}
\BIBentryALTinterwordspacing
A.~Mohan, H.~Sane, K.~Doshi, S.~Edupuganti, N.~Nayak, and V.~Sukhomlinov,
  ``Agile cold starts for scalable serverless,'' in \emph{11th {USENIX}
  Workshop on Hot Topics in Cloud Computing (HotCloud 19)}.\hskip 1em plus
  0.5em minus 0.4em\relax Renton, WA: {USENIX} Association, Jul. 2019.
  [Online]. Available:
  \url{https://www.usenix.org/conference/hotcloud19/presentation/mohan}
\BIBentrySTDinterwordspacing

\bibitem{vhive}
\BIBentryALTinterwordspacing
D.~Ustiugov, P.~Petrov, M.~Kogias, E.~Bugnion, and B.~Grot, ``Benchmarking,
  analysis, and optimization of serverless function snapshots,'' in
  \emph{Proceedings of the 26th ACM International Conference on Architectural
  Support for Programming Languages and Operating Systems}, ser. ASPLOS
  2021.\hskip 1em plus 0.5em minus 0.4em\relax New York, NY, USA: Association
  for Computing Machinery, 2021, p. 559–572. [Online]. Available:
  \url{https://doi.org/10.1145/3445814.3446714}
\BIBentrySTDinterwordspacing

\bibitem{demystifying}
\BIBentryALTinterwordspacing
M.~Kiener, M.~Chadha, and M.~Gerndt, ``Towards demystifying intra-function
  parallelism in serverless computing,'' in \emph{Proceedings of the Seventh
  International Workshop on Serverless Computing (WoSC7) 2021}, ser. WoSC
  '21.\hskip 1em plus 0.5em minus 0.4em\relax New York, NY, USA: Association
  for Computing Machinery, 2021, p. 42–49. [Online]. Available:
  \url{https://doi.org/10.1145/3493651.3493672}
\BIBentrySTDinterwordspacing

\bibitem{tppfaas}
\BIBentryALTinterwordspacing
M.~Steinbach, A.~Jindal, M.~Chadha, M.~Gerndt, and S.~Benedict, ``Tppfaas:
  Modeling serverless functions invocations via temporal point processes,''
  \emph{IEEE Access}, vol.~10, pp. 9059--9084, 2022. [Online]. Available:
  \url{https://doi.org/10.1109/ACCESS.2022.3144078}
\BIBentrySTDinterwordspacing

\bibitem{slam}
\BIBentryALTinterwordspacing
G.~Safaryan, A.~Jindal, M.~Chadha, and M.~Gerndt, ``Slam: Slo-aware memory
  optimization for serverless applications,'' in \emph{2022 IEEE 15th
  International Conference on Cloud Computing (CLOUD)}, 2022, pp. 30--39.
  [Online]. Available: \url{https://doi.org/10.1109/CLOUD55607.2022.00019}
\BIBentrySTDinterwordspacing

\bibitem{Fox2017}
\BIBentryALTinterwordspacing
G.~C. Fox, V.~Ishakian, V.~Muthusamy, and A.~Slominski, ``{Status of Serverless
  Computing and Function-as-a-Service(FaaS) in Industry and Research},'' aug
  2017. [Online]. Available: \url{http://arxiv.org/abs/1708.08028}
\BIBentrySTDinterwordspacing

\bibitem{Wang2019a}
\BIBentryALTinterwordspacing
H.~Wang, D.~Niu, and B.~Li, ``{Distributed Machine Learning with a Serverless
  Architecture},'' in \emph{Proceedings - IEEE INFOCOM}, vol. 2019-April.\hskip
  1em plus 0.5em minus 0.4em\relax Institute of Electrical and Electronics
  Engineers Inc., apr 2019, pp. 1288--1296. [Online]. Available:
  \url{https://doi.org/10.1109/INFOCOM.2019.8737391}
\BIBentrySTDinterwordspacing

\bibitem{Jiang2021}
\BIBentryALTinterwordspacing
J.~Jiang, S.~Gan, Y.~Liu, F.~Wang, G.~Alonso, A.~Klimovic, A.~Singla, W.~Wu,
  and C.~Zhang, ``{Towards Demystifying Serverless Machine Learning
  Training},'' vol.~15, no.~21, may 2021. [Online]. Available:
  \url{http://dx.doi.org/10.1145/3448016.3459240}
\BIBentrySTDinterwordspacing

\bibitem{mlless}
\BIBentryALTinterwordspacing
M.~S\'{a}nchez-Artigas and P.~G. Sarroca, ``Experience paper: Towards enhancing
  cost efficiency in serverless machine learning training,'' in
  \emph{Proceedings of the 22nd International Middleware Conference}, ser.
  Middleware '21.\hskip 1em plus 0.5em minus 0.4em\relax New York, NY, USA:
  Association for Computing Machinery, 2021, p. 210–222. [Online]. Available:
  \url{https://doi.org/10.1145/3464298.3494884}
\BIBentrySTDinterwordspacing

\bibitem{ali2022smlt}
A.~Ali, S.~Zawad, P.~Aditya, I.~E. Akkus, R.~Chen, and F.~Yan, ``Smlt: A
  serverless framework for scalable and adaptive machine learning design and
  training,'' \emph{arXiv preprint arXiv:2205.01853}, 2022.

\bibitem{tensorflow}
\BIBentryALTinterwordspacing
M.~Abadi, P.~Barham, J.~Chen, Z.~Chen, A.~Davis, J.~Dean, M.~Devin,
  S.~Ghemawat, G.~Irving, M.~Isard, M.~Kudlur, J.~Levenberg, R.~Monga,
  S.~Moore, D.~G. Murray, B.~Steiner, P.~Tucker, V.~Vasudevan, P.~Warden,
  M.~Wicke, Y.~Yu, and X.~Zheng, ``{TensorFlow: A system for large-scale
  machine learning},'' in \emph{Proceedings of the 12th USENIX Symposium on
  Operating Systems Design and Implementation, OSDI 2016}, vol.~10, no. July,
  2016, pp. 265--283. [Online]. Available:
  \url{https://www.usenix.org/system/files/conference/osdi16/osdi16-liu.pdf}
\BIBentrySTDinterwordspacing

\bibitem{aws-cognito}
\BIBentryALTinterwordspacing
AWS, ``{Amazon Cognito - Simple and Secure User Sign Up {\&} Sign In | Amazon
  Web Services (AWS)},'' 2020. [Online]. Available:
  \url{https://aws.amazon.com/cognito/}
\BIBentrySTDinterwordspacing

\bibitem{Mothukuri2021}
V.~Mothukuri, R.~M. Parizi, S.~Pouriyeh, Y.~Huang, A.~Dehghantanha, and
  G.~Srivastava, ``{A survey on security and privacy of federated learning},''
  \emph{Future Generation Computer Systems}, vol. 115, pp. 619--640, feb 2021.

\bibitem{Kim2021}
\BIBentryALTinterwordspacing
M.~Kim, O.~G{\"{u}}nl{\"{u}}, and R.~F. Schaefer, ``{Federated Learning with
  Local Differential Privacy: Trade-offs between Privacy, Utility, and
  Communication},'' feb 2021. [Online]. Available:
  \url{http://arxiv.org/abs/2102.04737}
\BIBentrySTDinterwordspacing

\bibitem{Wei2020}
K.~Wei, J.~Li, M.~Ding, C.~Ma, H.~H. Yang, F.~Farokhi, S.~Jin, T.~Q. Quek, and
  H.~{Vincent Poor}, ``{Federated Learning with Differential Privacy:
  Algorithms and Performance Analysis},'' \emph{IEEE Transactions on
  Information Forensics and Security}, vol.~15, pp. 3454--3469, 2020.

\bibitem{ray}
\BIBentryALTinterwordspacing
P.~Moritz, R.~Nishihara, S.~Wang, A.~Tumanov, R.~Liaw, E.~Liang, M.~Elibol,
  Z.~Yang, W.~Paul, M.~I. Jordan, and I.~Stoica, ``Ray: A distributed framework
  for emerging {AI} applications,'' in \emph{13th USENIX Symposium on Operating
  Systems Design and Implementation (OSDI 18)}.\hskip 1em plus 0.5em minus
  0.4em\relax Carlsbad, CA: USENIX Association, Oct. 2018, pp. 561--577.
  [Online]. Available:
  \url{https://www.usenix.org/conference/osdi18/presentation/moritz}
\BIBentrySTDinterwordspacing

\bibitem{serverless_reliability}
R.~Xie, Q.~Tang, S.~Qiao, H.~Zhu, F.~R. Yu, and T.~Huang, ``When serverless
  computing meets edge computing: Architecture, challenges, and open issues,''
  \emph{IEEE Wireless Communications}, vol.~28, no.~5, pp. 126--133, 2021.

\bibitem{serverless_reliability_2}
S.~G. Kulkarni, G.~Liu, K.~K. Ramakrishnan, and T.~Wood, ``Living on the edge:
  Serverless computing and the cost of failure resiliency,'' in \emph{2019 IEEE
  International Symposium on Local and Metropolitan Area Networks (LANMAN)},
  2019, pp. 1--6.

\bibitem{gcf_sla}
\BIBentryALTinterwordspacing
G.~Cloud, ``Cloud functions service level agreement (sla),'' 2022. [Online].
  Available: \url{https://cloud.google.com/functions/sla}
\BIBentrySTDinterwordspacing

\bibitem{gcp_compute}
\BIBentryALTinterwordspacing
------, ``Compute engine service level agreement (sla),'' 2022. [Online].
  Available: \url{https://cloud.google.com/compute/sla}
\BIBentrySTDinterwordspacing

\bibitem{behind}
D.~{Kelly}, F.~{Glavin}, and E.~{Barrett}, ``Serverless computing: Behind the
  scenes of major platforms,'' in \emph{2020 IEEE 13th International Conference
  on Cloud Computing (CLOUD)}, 2020, pp. 304--312.

\bibitem{mohanty2018evaluation}
S.~K. Mohanty, G.~Premsankar, M.~Di~Francesco \emph{et~al.}, ``An evaluation of
  open source serverless computing frameworks.'' in \emph{CloudCom}, 2018, pp.
  115--120.

\bibitem{k8s}
\BIBentryALTinterwordspacing
``{Kubernetes},'' 2022. [Online]. Available: \url{https://kubernetes.io/}
\BIBentrySTDinterwordspacing

\bibitem{openwhisk}
\BIBentryALTinterwordspacing
{The Apache Foundation}, ``{Apache OpenWhisk is a serverless, open source cloud
  platform},'' 2018. [Online]. Available: \url{https://openwhisk.apache.org/}
\BIBentrySTDinterwordspacing

\bibitem{jonas2017occupy}
E.~Jonas, Q.~Pu, S.~Venkataraman, I.~Stoica, and B.~Recht, ``Occupy the cloud:
  Distributed computing for the 99\%,'' in \emph{Proceedings of the 2017
  Symposium on Cloud Computing}, 2017, pp. 445--451.

\bibitem{ema}
F.~Klinker, ``Exponential moving average versus moving exponential average,''
  \emph{Mathematische Semesterberichte}, vol.~58, no.~1, pp. 97--107, 2011.

\bibitem{dbscan}
M.~Ester, H.-P. Kriegel, J.~Sander, X.~Xu \emph{et~al.}, ``A density-based
  algorithm for discovering clusters in large spatial databases with noise.''
  in \emph{kdd}, vol.~96, no.~34, 1996, pp. 226--231.

\bibitem{ch-index}
\BIBentryALTinterwordspacing
T.Caliński and J.~Harabasz, ``A dendrite method for cluster analysis,''
  \emph{Communications in Statistics}, vol.~3, no.~1, pp. 1--27, 1974.
  [Online]. Available:
  \url{https://www.tandfonline.com/doi/abs/10.1080/03610927408827101}
\BIBentrySTDinterwordspacing

\bibitem{8758926}
A.~V. Papadopoulos, L.~Versluis, A.~Bauer, N.~Herbst, J.~v. Kistowski,
  A.~Ali-Eldin, C.~L. Abad, J.~N. Amaral, P.~Tůma, and A.~Iosup,
  ``Methodological principles for reproducible performance evaluation in cloud
  computing,'' \emph{IEEE Transactions on Software Engineering}, vol.~47,
  no.~8, pp. 1528--1543, 2021.

\bibitem{thesismohamed}
\BIBentryALTinterwordspacing
M.~Elzohairy, ``Mitigation of stragglers in serverless federated learning,''
  2022. [Online]. Available:
  \url{https://mediatum.ub.tum.de/doc/1685641/1685641.pdf}
\BIBentrySTDinterwordspacing

\bibitem{lecun-mnist}
\BIBentryALTinterwordspacing
Y.~LeCun and C.~Cortes, ``{MNIST handwritten digit database},'' 2010. [Online].
  Available: \url{http://yann.lecun.com/exdb/mnist/}
\BIBentrySTDinterwordspacing

\bibitem{cohen2017emnist}
\BIBentryALTinterwordspacing
G.~Cohen, S.~Afshar, J.~Tapson, and A.~van Schaik, ``{EMNIST: an extension of
  MNIST to handwritten letters},'' feb 2017. [Online]. Available:
  \url{http://arxiv.org/abs/1702.05373}
\BIBentrySTDinterwordspacing

\bibitem{shakespeare-gutenberg}
\BIBentryALTinterwordspacing
``{The Complete Works of William Shakespeare by William Shakespeare - Free
  Ebook}.'' [Online]. Available: \url{http://www.gutenberg.org/ebooks/100}
\BIBentrySTDinterwordspacing

\bibitem{caldas2018leaf}
\BIBentryALTinterwordspacing
S.~Caldas, S.~M.~K. Duddu, P.~Wu, T.~Li, J.~Kone{\v{c}}n{\'{y}}, H.~B. McMahan,
  V.~Smith, and A.~Talwalkar, ``{LEAF: A Benchmark for Federated Settings},''
  in \emph{Workshop on Federated Learning for Data Privacy and Confidentiality,
  NeurIPS}, 2018, pp. 1--9. [Online]. Available:
  \url{http://arxiv.org/abs/1812.01097}
\BIBentrySTDinterwordspacing

\bibitem{fedscale}
F.~Lai, Y.~Dai, X.~Zhu, H.~V. Madhyastha, and M.~Chowdhury, ``Fedscale:
  Benchmarking model and system performance of federated learning,'' in
  \emph{Proceedings of the First Workshop on Systems Challenges in Reliable and
  Secure Federated Learning}, 2021, pp. 1--3.

\bibitem{hochreiter1997long}
S.~Hochreiter and J.~Schmidhuber, ``{Long short-term memory},'' \emph{Neural
  computation}, vol.~9, no.~8, pp. 1735--1780, 1997.

\bibitem{adam}
D.~P. Kingma and J.~Ba, ``Adam: A method for stochastic optimization,''
  \emph{arXiv preprint arXiv:1412.6980}, 2014.

\bibitem{sgd}
S.~Ruder, ``An overview of gradient descent optimization algorithms,''
  \emph{arXiv preprint arXiv:1609.04747}, 2016.

\bibitem{lrzcc}
{LRZ Compute Cloud}, \url{https://doku.lrz.de/display/PUBLIC/Compute+Cloud},
  accessed on 09/24/2020.

\bibitem{nginx}
\BIBentryALTinterwordspacing
``{NGINX | High Performance Load Balancer, Web Server, {\&} Reverse Proxy}.''
  [Online]. Available: \url{https://www.nginx.com/}
\BIBentrySTDinterwordspacing

\bibitem{knative}
{Knative}, \url{https://knative.dev/docs/}, accessed 09/24/2020.

\bibitem{li2020learning}
S.~Li, Y.~Cheng, W.~Wang, Y.~Liu, and T.~Chen, ``Learning to detect malicious
  clients for robust federated learning,'' \emph{arXiv preprint
  arXiv:2002.00211}, 2020.

\bibitem{zhang2022fldetector}
Z.~Zhang, X.~Cao, J.~Jia, and N.~Zhenqiang~Gong, ``Fldetector: Defending
  federated learning against model poisoning attacks via detecting malicious
  clients,'' \emph{arXiv e-prints}, pp. arXiv--2207, 2022.

\bibitem{fl_challenges}
P.~M. Mammen, ``Federated learning: Opportunities and challenges,'' \emph{arXiv
  preprint arXiv:2101.05428}, 2021.

\bibitem{fl_data_poisoning}
V.~Tolpegin, S.~Truex, M.~E. Gursoy, and L.~Liu, ``Data poisoning attacks
  against federated learning systems,'' in \emph{European Symposium on Research
  in Computer Security}.\hskip 1em plus 0.5em minus 0.4em\relax Springer, 2020,
  pp. 480--501.

\bibitem{gcf_cost}
\BIBentryALTinterwordspacing
G.~Cloud, ``Cloud functions pricing,'' 2022. [Online]. Available:
  \url{https://cloud.google.com/functions/pricing}
\BIBentrySTDinterwordspacing

\end{thebibliography}


\end{document}